\documentclass[aps,prb,twocolumn,superscriptaddress,raggedbottom,showpacs,floatfix]{revtex4}
\usepackage{graphicx}
\usepackage{bm}
\usepackage{amsmath}
\usepackage{amssymb}
\usepackage{amsthm}
\usepackage{amsfonts}
\usepackage{enumerate}
\usepackage{latexsym}
\usepackage{hyperref}
\usepackage{subfigure}

\newcommand {\be}{\begin{eqnarray}}
\newcommand {\ee}{\end{eqnarray}}
\newcommand {\rmd} {{\rm d}}

\newcommand{\bk}{{\bf k}}

\newcommand{\bR}{{\bf R}}

\newcommand{\bS}{{\bf S}}

\newcommand{\fvec}[1]{\boldsymbol{#1}}

\begin{document}

\title{Transport Theory of Metallic B20 Helimagnets}
\author{Jian Kang}
\email{jkang@umn.edu} \affiliation{School of Physics and Astronomy,
University of Minnesota, Minneapolis, MN 55455, USA}
\author{Jiadong Zang}
\email{jiadongzang@gmail.com} \affiliation{Department of Physics and
Astronomy, Johns Hopkins University, Baltimore, MD 21218, USA}
\date{\today }
\begin{abstract}
B20 compounds are a class of cubic helimagnets harboring nontrivial
spin textures such as spin helices and skyrmions. It has been well
understood that the Dzyaloshinskii-Moriya (DM) interaction is the
origin of these textures, and the physics behind the DM interaction
is the spin-orbital coupling (SOC). However the SOC shows its effect
not only on the spins, but also on the electrons. In this paper, we
will discuss effects of the SOC on the electron and spin transports
in B20 compounds. An effective Hamiltonian is presented from
symmetry analysis, and the spin-orbital coupling therein shows
anomalous behaviors in anisotropic magnetoresistance (AMR) and
helical resistance. New effects such as inverse spin-galvanic effect
is proposed, and the origin of the DM interaction is discussed.
\end{abstract}
 \maketitle

\section{Introduction}

Symmetry is a central topic of modern physics and material science.
Reduced symmetry has given rise to innumerable novel phenomena. For
example, breaking of the translational symmetry leads to the
emergence of lattices and crystals, which are the platforms of
condensed matter studies. The highest symmetry of a lattice has the
point group of $O_h$, where most of the ferromagnetic materials
belong to. Surprises have been brought by further reducing this
symmetry. B20 compound, with representatives of
FeSi\cite{schlesinger_unconventional_1993,ditusa_metal-insulator_1997},
MnSi\cite{pfleiderer_partial_2004,muhlbauer_skyrmion_2009,ritz_formation_2013}
, FeGe\cite{yu_near_2011,huang_extended_2012}, and
Cu$_2$SeO$_3$\cite{seki_observation_2012}, is such an interesting
class of materials. Although B20 compound has a cubic lattice, it
has the lowest symmetry in this crystal system. Complicated
distributions of atoms dramatically bring down the symmetry, where
inversion, mirror, or four-fold rotational symmetries are absent.
Abundant phenomena are emerging consequently, among which the most
attractive one is the presence of nontrivial spin textures like
helices and skyrmions.

Spin helix is a spatially modulated magnetic texture. It is present
in magnetic materials with competing interactions. It has been
observed in B20 family Fe$_{1-x}$Co$_x$Si by real-space
imaging\cite{uchida_real-space_2006}. In this case, the
Dzyaloshinskii-Moriya (DM)
interaction\cite{dzyaloshinsky_thermodynamic_1958,moriya_anisotropic_1960}
\begin{equation}
H_{\text{DM}}=\mathbf{D}\cdot(\bS_i\times\bS_j)\label{eq:DMinteraction}
\end{equation}
between neighboring spins plays an important role. Broken inversion
symmetry in B20 compounds is the physical origin of this
interaction. Under an inversion operation about the center of the
joint line, two neighboring spins are exchanged, and the DM
interaction flips sign due to its cross product nature. In contrast,
the Heisenberg exchange, $H_\text{H}=-J\bS_i\cdot\bS_j$ is unchanged
under this operation, thus respects the inversion symmetry. The
Heisenberg exchange tends to align neighboring spins, while the DM
interaction tends to form an angle of $\pi/2$. As a result of the
competition, a finite angle is expanded by these two spins, whose
successive arrangement generates the spin helix. Spin helix is not
the only result of breaking inversion symmetry, but also the
magnetic
skyrmion\cite{skyrme_non-linear_1961,rossler_spontaneous_2006,muhlbauer_skyrmion_2009,yu_real-space_2010},
a topological spin texture. It is stabilized in B20 compounds at
finite magnetic fields and temperatures.

In the light of nontrivial spin modulation in helix and promising
spintronics applications of the skyrmion, a throughout understanding
of the electron and spin transports in B20 compounds is an urgent
subject. Longitudinal magnetoresistance measurements have been
performed to map out the phase diagrams containing helical and
skyrmion phases\cite{kadowaki_magnetization_1982,du_highly_2014}. On
the other hand, due to the emergent
electromagnetism\cite{zang_dynamics_2011,schulz_emergent_2012},
skyrmion phase can be precisely determined by the Hall
measurements\cite{neubauer_topological_2009,lee_unusual_2009,kanazawa_large_2011,huang_extended_2012,li_robust_2013}.
A peculiar non-Fermi liquid behavior is also addressed in MnSi
single crystals\cite{pfleiderer_partial_2004,ritz_formation_2013},
which is intimately related to the topology of spin
textures\cite{watanabe_anomalous_2014}. However, a deep study of the
spin transports in B20 compounds still lacks. Recently, an
experiment on anisotropic magnetoresistance (AMR) is performed in
bulk samples of
Fe$_{1-x}$Co$_x$Si\cite{huang_magnetoresistance_2014}. It is
surprisingly observed that compared to usual AMR in cobalt or other
cubic ferromagnetic materials, the magnetoresistance shows two,
instead of four, peaks. It shows that the system lacks of the
four-fold rotational symmetry, which is compatible with the reduced
symmetry in B20 compounds. However the microscopic origin waits to
be revealed. In another experiment, the measurement on helical
resistance shows an ultra-low resistance ratio of $1.35$ with
current parallel and perpendicular to the
helix\cite{huang_universal_2014}. It has predicted theoretically and
well tested experimentally that this ratio should be larger than
$3$. This observation apparently violate this common concept.

These two experiments suggest a new mechanism involving nontrivial
spin scatterings, and thus inevitably call for the important effects
of the spin-orbital coupling (SOC) on the conduction electrons in
B20 compounds. The SOC has already shows its power in the spin
interactions in B20 compounds. It is well known that a nonvanishing
DM interaction requires not only the inversion symmetry breaking,
but also a large SOC\cite{moriya_anisotropic_1960}. However effects
of the SOC on conduction electrons have never been discussed. In
this paper, we will show that the SOC well explains the two
experiments above, and provides several other proposals.

This paper is organized as follows. In the following section, an
effective Hamiltonian is constructed, where both linear and cubic
SOC terms are present. Section III shows that these SOC terms are
the microscopic origin of the DM interactions in B20 compounds. In
section IV and V, the linear SOC gives rise to inverse spin-galvanic
effect and ultra-low ratio of helical resistance, respectively. In
section VI, the importance of the cubic SOC is revealed, which
provides the microscopic mechanism of the anomalous AMR effect.

\section{Effective Hamiltonian}

In order to understand the transport behaviors of B20 compounds, the
first priority is to construct the effective Hamiltonian of the
conduction electrons. As we are interested in the long-range
behaviors, only momenta around the $\Gamma$ point will be relevant.
The bands around other high symmetry points in the Brillouin zone
might cross the Fermi energy, and contribute to the transports in
some form. However, the qualitative behavior, especially the
symmetries, will not change.

The importance of the SOC indicates that the conventional quadratic
dispersion, $H(\bk)=\hbar^2\bk^2/2m$, is not adequate in
understanding the spin-related transports. Therefore, additional
terms coupling spin and momentum is called for. To this end, we
analyze the symmetry of the B20 compounds and employ the theory of
invariants to construct the effective
Hamiltonian\cite{winkler_spin--orbit_2003}.

In the international notation, the space group of B20 compounds is
$P_{2_13}$, where $2$ and $3$ mean the two-fold and three-fold
rotational symmetries respectively, while $1$ indicates a fractal
translation in the space group operation. No other symmetries are
present. Around the $\Gamma$ point, the fractal rotation does not
change the space group irreducible
representations\cite{dresselhaus_group_2010}, thus is not relevant,
and the space group is isomorphic to its K-group, which is a group
containing all the point group operations within the space group.
Irreducible representations of the space group at $\Gamma$ point is
the same as that of the K-group. For $P_{2_13}$, the K-group is the
$T23$ group, containing 12 elements; 1 identity, 3 $C_2$ rotations
($\pi$ rotations about axis [100], [010], and [001], and 8 $C_3$
rotations (Clockwise rotations of $2\pi/3$ about directions
[$\pm$1,$\pm$1,$\pm$1]). Compared to the complete group operations
in the $O_h$ group, four-fold rotations, inversions, and mirror
symmetries are absent.

Once spin is taken into account, a $2\pi$ rotation in the spin space
reverses the sign, and should be treated as an additional group
operation. One thus has to consider the double group of $T23$, whose
irreducible representations and characters are listed in Tab.
(\ref{tab:Character2}). Operations with bars are joint action of
point group element and the $2\pi$ spin rotation.

\begin{table}[h]
  \centering
      \caption{ The character table for the double group of $T23$. }
\label{tab:Character2} \hspace{0cm}
\begin{tabular}{cccccccc}
    \hline\hline
    $T23$  & $E$ & $\bar{E}$ & $4C_3$ & $4\bar{C}_3$ & $3C_2+3\bar{C}_2$ & $4C_{3}^{-1}$
    & $4\bar{C}_3^{-1}$\\
    \hline
   $\Gamma_1$ &  1 &  1 &  1 &  1 &  1 &  1 &  1\\
    $\Gamma_2$ &  1 &  1 &  $w$ &  $w$ &  1 & $w^2$ &  $w^2$\\
    $\Gamma_3$ &  1 &  1 &  $w^2$ &  $w^2$ & 1 &  $w$ &  $w$\\
    $\Gamma_4$ &  3 &  3 &  0 &  0 & -1 &  0 & 0\\
    $\Gamma_5$ &  2 & -2 &  1 &  -1 & 0 & 1 & -1\\
    $\Gamma_6$ &  2 & -2 &  $w$ &  $-w$ &  0 &  $w^2$ & $-w^2$\\
    $\Gamma_7$ &  2 & -2 &  $w^2$ & $-w^2$ & 0 & $w$ &  $-w$\\
   \hline
  \end{tabular}%
\end{table}

First principle calculations on metallic B20 materials, such as MnSi
and FeGe, show that the orbitals around the Fermi surface are mainly
d-orbitals\cite{jeong_implications_2004}. Under spin-orbit coupling,
these orbitals are split into $j=5/2$ and $j=3/2$ orbitals, which
correspond to $D_{5/2}$ and $D_{3/2}$ irreducible representations of
the rotation group respectively. The character of a rotation of angle $%
\alpha $ in $D_j$ is given by
\begin{equation}
\chi=\frac{\sin((j+\frac{1}{2})\alpha)}{\sin(\frac{1}{2}\alpha)}
\end{equation}
In the crystal field, these orbitals are further split into
suborbitals, which corresponds to the decomposition in terms of
$T23$'s irreducible representations;
\begin{equation}
D_{5/2}\downarrow T=\Gamma_5\oplus\Gamma_6\oplus\Gamma_7
\end{equation}
\begin{equation}
D_{3/2}\downarrow T=\Gamma_6\oplus\Gamma_7
\end{equation}
In reality, all these suborbitals might be relevant around the Fermi
surface. In addition, there are four magnetic atoms in each unit
cell, leading to a total of 20 bands. In order to capture the key
feature of these materials, we would like to keep the Hamiltonian in
the minimal form. Only one band out of three representations
$\Gamma_{5,6,7}$ will be considered. The resulting Hamiltonian will
be two by two, the simplest Hamiltonian taking into account the SOC.

The model Hamiltonian $H$ communicating the Hilbert spaces corresponding to irreducible representations $%
\alpha $ and $\beta $ is in general%
\begin{equation}
H(\mathbf{k})=\sum_{l_{1}^{\alpha }=1}^{n_{\alpha
}}\sum_{l_{2}^{\beta }=1}^{n_{\beta }}h(\mathbf{k})|l_{1}^{\alpha
}\rangle \langle l_{2}^{\beta }|
\end{equation}%
Here $n_{\alpha }$ and $n_{\beta }$ are dimensions of these two
representations respectively. In the current case, band mixing is
neglected such that $\alpha =\beta $.
$l_{1}^{\alpha }$ labels the basis in the irreducible representation $\alpha $. The operator part $%
|l_{1}^{\alpha }\rangle \langle l_{2}^{\beta }|$ transforms as the
product representation $\Gamma _{\alpha }^{\ast }\times \Gamma
_{\beta }$, which can
be decomposed as the direct sum of irreducible representations $%
\Gamma _{\alpha }^{\ast }\times \Gamma _{\beta }=\oplus _{\gamma
}\Gamma _{\gamma }$. The basis $X_{l_{3}^{\gamma }}^{\gamma }$ of an
irreducible representation $\Gamma _{\gamma }$ contained in this
product representation is the superposition of direct products
$|l_{1}^{\alpha }\rangle \langle
l_{2}^{\beta }|$ as $%
X_{l_{3}^{\gamma }}^{\gamma }=\sum_{l_{1}^{\alpha }=1}^{n_{\alpha
}}\sum_{l_{2}^{\beta }=1}^{n_{\beta }}C_{l_{1}^{\alpha }l_{2}^{\beta
},l_{3}^{\gamma }}^{\alpha \beta ,\gamma }|l_{1}^{\alpha }\rangle
\langle l_{2}^{\beta }|$, where $C_{l_{1}^{\alpha }l_{2}^{\beta
},l_{3}^{\gamma }}^{\alpha \beta ,\gamma }$ are the Clebsh-Gordan coefficients\cite{koster_space_1957}. To keep the Hamiltonian invariant under group operations, $%
h(\mathbf{k})$ must be an irreducible tensor operator in the representation $%
\Gamma _{\gamma }^{\ast }$, such that $H(\mathbf{k})$ belongs to the
trivial representation $\Gamma _{1}$ contained in the product
representation $\Gamma _{\gamma }^{\ast }\times \Gamma _{\gamma }$.
As a consequence, the invariant Hamiltonian is given by
\begin{eqnarray}
H(\mathbf{k}) &=&\sum_{\gamma }^{\alpha \beta }a_{\gamma }^{\alpha
\beta
}\sum_{1}^{n_{\gamma }}h_{l_{3}^{\gamma }}^{\gamma }(\mathbf{k}%
)X_{l_{3}^{\gamma }}^{\gamma } \\
&=&\sum_{\gamma }^{\alpha \beta }a_{\gamma }^{\alpha \beta
}\sum_{1}^{n_{\gamma }}h_{l_{3}^{\gamma }}^{\gamma
}(\mathbf{k})\left( \sum_{l_{1}^{\alpha }=1}^{n_{\alpha
}}\sum_{l_{2}^{\beta }=1}^{n_{\beta }}C_{l_{1}^{\alpha }l_{2}^{\beta
},l_{3}^{\gamma }}^{\alpha \beta ,\gamma
}|l_{1}^{\alpha }\rangle \langle l_{2}^{\beta }|\right)   \notag \\
&&\label{eq:HamiltonianGroup}
\end{eqnarray}
where coefficients $a_{\gamma }^{\alpha \beta }$ are free parameters
that cannot be dictated from the symmetry analysis.

For $\Gamma _{5}$, $\Gamma _{5}^{\ast }\times \Gamma _{5}=\Gamma
_{1}\oplus
\Gamma _{4}$. However $X_{1}^{\Gamma _{1}}$ and all the three matrices $%
X_{1,2,3}^{\Gamma _{4}}$ are all trivial identity matrices.
Therefore the effective can be reduced to be spinless, and
spin-orbital coupling is absent. The corresponding Hamiltonian is
therefore the simplest quadratic one $H(\bk)=\hbar^2\bk^2/2m$, which
does not bring anything new. New physics comes when we turn to
$\Gamma _{6}$ or $\Gamma _{7}$ representations. As these two
representations are complex conjugate to each other, the effective
Hamiltonians are the same. In the following, we will take $\Gamma
_{6}$ for
example without loss of generality.%
\begin{equation}
\Gamma _{6}^{\ast }\times \Gamma _{6} =\Gamma _{1}\oplus \Gamma _{4}
\end{equation}
\begin{eqnarray}
X_{1}^{\Gamma _{1}} &=&\left(
\begin{array}{cc}
\text{ \ }1\text{ \ } & 0 \\
0 & \text{ \ }1\text{ \ }%
\end{array}%
\right) \text{, }X_{1}^{\Gamma _{4}}=\left(
\begin{array}{cc}
0 & -i/\sqrt{3} \\
-i/\sqrt{3} & 0%
\end{array}%
\right)   \notag \\
X_{2}^{\Gamma _{4}} &=&\left(
\begin{array}{cc}
0 & 1/\sqrt{3} \\
-1/\sqrt{3} & 0%
\end{array}%
\right) \text{, }X_{3}^{\Gamma _{4}}=\left(
\begin{array}{cc}
-i/\sqrt{3} & 0 \\
0 & i/\sqrt{3}%
\end{array}%
\right)   \notag \\
&&
\end{eqnarray}%
It shows explicitly that $X_{i}^{\Gamma _{4}}=-i/\sqrt{3}\sigma
_{i}\,$, where $\sigma _{i}$ are three Pauli matrices. The overall
factor -$i/\sqrt{3} $ can be absorbed into the factor $a_{\gamma
}^{\alpha \beta }$ in Eq. (\ref{eq:HamiltonianGroup}). Now the
remaining job is to construct the irreducible tensor operators
$h_{l_{3}^{\gamma }}^{\gamma }(\mathbf{k})$. For $\Gamma
_{1}$, it is very simple that $h_{1}^{\Gamma _{1}}(\mathbf{k})=\mathbf{k}%
^{2}/2m$. While for $\Gamma _{4}$, two basis may apply. $h_{i}^{\Gamma _{4}}(\mathbf{k}%
)=(k_{x},k_{y},k_{z})$, or
$h_{i}^{\Gamma_{4}}=(k_{x}(k_{y}^{2}-k_{z}^{2}),k_{y}(k_{z}^{2}-k_{x}^{2}),k_{z}(k_{x}^{2}-k_{y}^{2}))$,
which are on the first and third orders in momentum $\bk$
respectively. Terms of second order in $\bk$ break the time reversal
symmetry once coupled to the spin, thus can be neglected. As a
consequence, the effective Hamiltonian for conduction electrons in
B20 compounds is given by
\begin{eqnarray}
H &=&\frac{\mathbf{k}^{2}}{2m}+\alpha (k_{x}\sigma _{x}+k_{y}\sigma
_{y}+k_{z}\sigma _{z})\notag \\
&+&\beta \lbrack k_{x}\sigma _{x}(k_{y}^{2}-k_{z}^{2})+k_{y}\sigma
_{y}(k_{z}^{2}-k_{x}^{2})+k_{z}\sigma _{z}(k_{x}^{2}-k_{y}^{2})]
\notag\\
&& \label{fullHamiltonian}
\end{eqnarray}%

This Hamiltonian can also be intuitively guessed from simple
symmetry analysis. The presence of $C_3$ symmetry enforces the
permutation symmetry in the Hamiltonian, while the $C_2$ symmetry
rules out most of the combinations. It is worth emphasizing that the
linear spin-orbital coupling $\bk\cdot\boldsymbol{\sigma}$ is not
adequate, as it is a full rotational symmetric term. $C_4$ symmetry
is also respected by this term, but is apparently broken in $T23$
group. It is the cubic spin-orbital coupling, the last term in Eq.
(\ref{fullHamiltonian}), that breaks $C_4$. Therefore Eq.
(\ref{fullHamiltonian}) is the minimal Hamiltonian that faithfully
describes the symmetry of B20 compounds.

The cubic spin-orbital coupling is well known in III-V
semiconductors induced by bulk inversion asymmetry (BIA)
\cite{zutic_spintronics:_2004}. The linear term
$\bk\cdot\boldsymbol{\sigma}$ is a new term. It will be discussed in
the following that it captures most of the nontrivial physics in B20
compounds. As $\boldsymbol{\sigma}$ is a pseudo-vector,
$\bk\cdot\boldsymbol{\sigma}$ is a pseudo-scalar. This is thus a
forbidden term in lattices with any inversion or mirror symmetries.
That is why it is absent in the III-V semiconductors and most of the
ferromagnetic materials. However, elements in the $T23$ point group
are only pure rotations, so that $\bk\cdot\boldsymbol{\sigma}$ is
allowed, and contributes significantly to the long range behaviors.

\section{Origin of Spin Interactions}

Real space images of B20 compounds have shown that the spin helix
therein looks like a successive array of Bloch
walls\cite{uchida_real-space_2006}, where magnetizations are
rotating in a plane perpendicular to their propagation direction. In
addition, the skyrmion has a double twist
structure\cite{yu_real-space_2010}. These features are well
described by the DM interaction in the following shape
\begin{equation}
H_{DM}=D\hat{r}_{ij}\cdot(\bS_i\times\bS_j)\label{eq:DMB20},
\end{equation}
namely the DM vector $\mathbf{D}$ in Eq. (\ref{eq:DMinteraction})
should point from one spin to the other. Although it is compatible
with the symmetry \cite{leonov_chiral_2014}, the microscopic origin
still lacks. However, it can be understood by the SOC in our
effective Hamiltonian as follows.

Quantitatively, we can employ the field approach to calculate the
Ruderman-Kittel-Kasuya-Yosida (RKKY) interaction between two
neighboring spins $\bS_1$ and
$\bS_2$\cite{ruderman_indirect_1954,kasuya_theory_1956,yosida_magnetic_1957}.
The electron's action is given by
\begin{eqnarray}
S &=&\sum_{n}\int d^{3}\mathbf{k}\bar{\psi}(-\mathbf{k},-i\omega
_{n})(-i\omega _{n}+\frac{\mathbf{k}^{2}}{2m}+\alpha \mathbf{k\cdot \sigma }%
)\psi (\mathbf{k},i\omega _{n})  \nonumber \\
&&+\sum_{i=1}^{2}\sum_{n}\int d^{3}\mathbf{k}\bar{\psi}(-\mathbf{k-q}%
,-i\omega _{n})\mathbf{S}_{i}\cdot \mathbf{\sigma }\text{e}^{-i\mathbf{q}%
\cdot \mathbf{R}_{i}}\psi (\mathbf{k},i\omega _{n}) \notag\\
&&
\end{eqnarray}
where $\mathbf{R}_i$ are the positions of two spins. The spin
interaction can be derived by integrating out the electrons using
the gradient expansion. Up to the second order, the spin-spin
interaction is given by
\begin{equation}
S_{\text{eff}}=-2\sum_{n}\int d^{3}\mathbf{k}Tr[G(\bR,i\omega _{n})\mathbf{S}%
_{1}\cdot \mathbf{\sigma }G(-\bR,i\omega _{n})\mathbf{S}_{2}\cdot \mathbf{%
\sigma }]\label{eq:RKKYaction}
\end{equation}
where $\mathbf{R}=\mathbf{R}_1-\mathbf{R}_2$, and $G(\bR,i\omega
_{n})$ is the real-space Green's function defined as
\begin{eqnarray}
G(\bR,i\omega _{n})&=&\int
d^{3}\mathbf{k}\frac{\text{e}^{-i\mathbf{k}\cdot
\mathbf{R}}}{-i\omega _{n}+\frac{\mathbf{k}^{2}}{2m}-\alpha
\mathbf{k\cdot \sigma }} \\
&=&\int_{0}^{\infty }dk\int_{0}^{\pi }d\theta k^2\sin\theta\nonumber\\
&&\times\int_{0}^{2\pi }d\varphi \frac{%
i\omega _{n}-\frac{k^{2}}{2m}+\alpha \mathbf{k\cdot \sigma }}{(-i\omega _{n}+%
\frac{k^{2}}{2m})^{2}-\alpha ^{2}k^{2}}\text{e}^{-i\mathbf{k}\cdot \mathbf{R}%
} \notag\\
&&
\end{eqnarray}%
One can decompose the momentum $\mathbf{k}$ into directions parallel
with and perpendicular to $\hat{R}$ as
$\bk=(\bk\cdot\hat{R})\hat{R}+(\hat{R}\times\bk)\times\hat{R}\equiv\bk_\parallel+\bk_\perp$.
Apparently, because of $\bk_\perp\cdot\mathbf{R}=0$,
$\exp(i\bk_\perp\cdot\bR)=1$, and
\begin{equation}
\int_{0}^{2\pi }d\varphi \frac{\alpha \mathbf{k}_{\perp
}\mathbf{\cdot \sigma }}{(-i\omega _{n}+\frac{k^{2}}{2m})^{2}-\alpha
^{2}k^{2}}=0.
\end{equation}
The Only contribution comes from the coupling between
$\bk_\parallel$ and Pauli matrices. Thus
\begin{eqnarray}
G(\bR,i\omega _{n})&=&\int_{0}^{\infty }dk\int_{0}^{\pi }d\theta
k^{2}\sin \theta\nonumber\\
&&\times\int_{0}^{2\pi }d\varphi  \frac{i\omega _{n}-\frac{k^{2}}{2m}%
+\alpha \mathbf{k}_{\parallel }\mathbf{\cdot \sigma }}{(-i\omega _{n}+\frac{%
k^{2}}{2m})^{2}-\alpha ^{2}k^{2}}\text{e}^{-i\mathbf{k}_{\parallel
}\cdot
\mathbf{R}}  \nonumber \\
&=&2\pi \int_{0}^{\infty }dk\int_{0}^{\pi }d\theta k^{2}\sin \theta \nonumber\\
&&\times\frac{%
i\omega _{n}-\frac{k^{2}}{2m}+\alpha k\cos \theta
\hat{R}\mathbf{\cdot
\sigma }}{(-i\omega _{n}+\frac{k^{2}}{2m})^{2}-\alpha ^{2}k^{2}}\text{e}%
^{-ikR\cos \theta } \notag\\
&\equiv &G_{0}(R)+G_{1}(R)\bR\cdot\boldsymbol{\sigma }
\end{eqnarray}
where
\begin{equation}
G_{0}(R)=2\pi \int_{0}^{\infty }dk\int_{0}^{\pi }d\theta
\frac{k^{2}\sin
\theta (i\omega _{n}-\frac{k^{2}}{2m})}{(-i\omega _{n}+\frac{k^{2}}{2m}%
)^{2}-\alpha ^{2}k^{2}}\text{e}^{-ikR\cos \theta }
\end{equation}
and
\begin{equation}
G_{1}(R)=\frac{2\pi }{R}\int_{0}^{\infty }dk\int_{0}^{\pi }d\theta
 \frac{\alpha k^{3}\sin \theta\cos \theta }{(-i\omega _{n}+\frac{k^{2}}{2m}%
)^{2}-\alpha ^{2}k^{2}}\text{e}^{-ikR\cos \theta }
\end{equation}
One can easily get these Green's functions by evaluating contour
integrals. The real part of the poles of $k$ gives rise to the
Friedel oscillations.

Consequently, the effective RKKY Hamiltonian is given by
\begin{equation}
H^{\text{RKKY}}=-\frac{2}{\beta }\sum_{n}[(G_{0}^{2}+G_{1}^{2})\mathbf{S}%
_{1}\cdot \mathbf{S}_{2}+2G_{0}G_{1}\mathbf{R}\cdot
(\mathbf{S}_{1}\times \mathbf{S}_{2})]
\end{equation}
The first term in the Heisenberg exchange, while the second term is
the DM interaction. Compared to Eq. (\ref{eq:DMinteraction}), the DM
vector $\mathbf{D}$ is along $\bR=\bR_1-\bR_2$, which is consistent
with Eq. (\ref{eq:DMB20}) for B20 compounds. One can easily show
that even the cubic spin orbital coupling in Eq.
(\ref{fullHamiltonian}) is included, the direction of DM vector is
still unchanged. Thus the SOC in Eq. (\ref{fullHamiltonian}) gives
rise to the right DM interactions in B20, and therefore the right
physical origin. In the following two sections, we will discuss the
effects of these two SOC terms in the collective transports.

The physical picture behind this calculation is the following. By
the linear spin-orbital coupling
$\alpha\bk\cdot\boldsymbol{\sigma}$, the conduction electron feels
effectively a magnetic field $-\alpha\bk\sim-\alpha
m\mathbf{v}/\hbar$. Therefore once it hops from one site to the
other, its spin must process about the effective magnetic field
along the joint line between these two sites. The coupling between
conduction electron and local magnetic moments thus reduces the
energy once the moments at these two sites process in the same way.
The direction of the DM vector $\mathbf{D}$ is thus parallel with
the effective field, which point one spin to the other.

In reality, the interaction between the neighboring spins might have
various origins besides the RKKY mechanism. However the physical
picture of spin procession persists in any mechanism. Therefore the
DM interaction always has the desired form once the spin orbital
coupling, Eq. (\ref{fullHamiltonian}), is present.

Interesting consequence follows when an ultra-thin film of B20
compound is grown along [001] direction, and an electric field is
applied perpendicular to the film. The intrinsic linear SOC gives
$H=\alpha(k_x\sigma_x+k_y\sigma_y)$, while the additional Rashba SOC
induced by the electric field is
$H=\alpha_{R}(k_x\sigma_y-k_y\sigma_x)$. Still the intrinsic SOC
gives DM interactions with DM vector pointing from one spin to the
neighbor on the film. However the Rashba SOC contributes a DM vector
perpendicular to the intrinsic one. In B20 compounds the spin helix
looks like a successive array of Bloch domain walls, where the spins
are rotating in a plane perpendicular to the propagation direction.
However in the large $\alpha_R$ limit, the resulting spin helix is a
successive array of Neel walls, where the spins are coplanar to the
propagation direction. Therefore by increasing the electric field,
one can expect a gradual deformation of the spin helix. Similarly,
the skyrmion will deform to that generated by interfacial DM
interactions\cite{fert_skyrmions_2013}. These deformations can be
observed by Lorentz TEM images.

\section{Inverse Spin-Galvanic Effect}

There have been extensive discussions on the interaction between the
conduction electrons and local magnetic moments $\mathbf{M}$ in
metallic magnets. These two are directly coupled to each other via
the Hund's rule coupling $H=-J_H\mathbf{M}\cdot\boldsymbol{\sigma}$,
where $\boldsymbol{\sigma}$ is the spin of the conduction electron.
In the adiabatic limit, $J_H\rightarrow\infty$, electron spins are
parallel with the local moments. Algebraically, one can perform an
$SU(2)$ transformation $U$ such that
$U^\dagger\mathbf{M}\cdot\boldsymbol{\sigma}U=\sigma_z$. In the
adiabatic limit, only the up spin, namely the upper-left part of the
transformed Hamiltonian, is relevant. In case moments are spatially
non-uniform, the same $U$ rotation transforms the kinetic energy
$\bk^2/2m$ to $(\bk-e\mathcal{A})^2/2m$, where $SU(2)$ gauge field
$e\mathcal{A}_\mu=-iU^\dagger\partial_\mu
U$\cite{bazaliy_modification_1998}, whose upper-left part
$\mathbf{A}$ is the real-space emergent electromagnetic
field\cite{zang_dynamics_2011}. The minimal coupling between
$\mathbf{A}$ and the electric current $\mathbf{j}$ results in the
current driven domain wall or the skyrmion
motions\cite{tatara_theory_2004,zang_dynamics_2011}.

The scenario above is no longer valid in the presence of the SOC.
Due to the non-commutative nature of the Pauli matrices, the
conduction electrons feel more than the emergent electromagnetic
field $\mathbf{A}$, and additional coupling to the electric current
needs to be included. Careful analysis is required in the current
case.

One can similarly perform an $SU(2)$ gauge transformation $U$ to the
Hamiltonian
\begin{equation}
H=\frac{1}{2m}\mathbf{k}^{2}+\alpha
\mathbf{k}\cdot\boldsymbol{\sigma}
-J_{H}M(r)\cdot\boldsymbol{\sigma}
\end{equation}%
so that $U^{\dagger }M(r)\cdot\boldsymbol{\sigma} U=m\sigma _{z}$.
The first term again
gives rise to $U^{\dagger }(\mathbf{k}^{2}/2m)U=(k-e\mathcal{A})^{2}/2m$, with $%
e\mathcal{A}_{\mu }=-iU^{\dagger }\partial _{\mu }U$, while the
second one is
transformed as%
\begin{equation}
U^{\dagger }\mathbf{k}\cdot\boldsymbol{\sigma} U=U^{\dagger }\sigma
_{\mu }U(k_{\mu }-e\mathcal{A}_{\mu })
\end{equation}
In the adiabatic limit, we take the upper-left ($\uparrow\uparrow$)
part of the transformed Hamiltonian, so that
\begin{equation}
H=\frac{1}{2m}(\mathbf{k}-e\mathbf{A})^{2}+\alpha (\mathbf{f\cdot }%
\bk-g)-J_{H}m
\end{equation}%
where $f_{\mu }=[U^{\dagger }\sigma _{\mu }U]_{\uparrow \uparrow }$, and $%
g=[{\mu }U^{\dagger }\sigma _{\mu }Ue\mathcal{A}_\mu]_{\uparrow \uparrow }$.%
As $\mathbf{\dot{x}}=\frac{\partial H}{\partial \mathbf{k}}=\frac{1}{m}(%
\mathbf{k}-e\mathbf{A})+\alpha \mathbf{f}$, the Lagrangian is given by%
\begin{eqnarray}
L &=&\mathbf{k}\cdot \mathbf{\dot{x}}-H  \notag \\
&=&\frac{1}{2}m\mathbf{\dot{x}}^{2}+e\mathbf{A}\cdot (\mathbf{\dot{x}}%
-\alpha \mathbf{f})-m\alpha \mathbf{f}\cdot \mathbf{\dot{x}}+\frac{1}{2}%
m\alpha ^{2}\mathbf{f}^{2}+\alpha g\notag\\
&&
\end{eqnarray}
The electric current $\mathbf{j}=e\mathbf{\dot{x}}$ thus minimal couples to $%
(\mathbf{A}-\frac{m\alpha }{e}\mathbf{f})$, instead of $\mathbf{A}$
in the absence of the SOC. For local magnetizations $\hat{m}$=($\sin
\theta \cos \phi$, $\sin \theta \sin
\phi$, $\cos \theta$), let $U=u_{0}+i\mathbf{u}\cdot\boldsymbol{\sigma}$. Up to a gauge, we get $%
u_{0}=\frac{1}{2\sin (\theta /2)},\mathbf{u}=\frac{1}{2\sin (\theta /2)}\hat{%
z}\times \hat{m}$, and consequently $\mathbf{f}=\hat{m}$. The
effective coupling between magnetization and the current
is thus given by%
\begin{equation}
L_{c}=\frac{1}{e}\mathbf{j}\cdot (e\mathbf{A}-\alpha \mathbf{m})
\end{equation}%
By varying the total action, the magnetization dynamics obeys the
following equation of motion;
\begin{equation}
\mathbf{\dot{m}}+\frac{1}{e}\mathbf{j}\cdot \nabla \mathbf{m}-\frac{\alpha }{%
e}\mathbf{m}\times \mathbf{j}+\mathbf{m}\times \mathbf{H}_{eff}=0
\end{equation}%
The third term is the contribution from SOC. It shows explicitly
that via SOC, an electric current $\mathbf{j}$ serves as an
effective planar field acting on the magnetizations, manifesting the
inverse spin-galvanic effect.

The physics of the inverse spin-galvanic effect is very simple.
Under a steady electric current, the Fermi surface acquires a shift
along the current direction, and ends up with a nonvanishing average
momentum $<\bk>$ in the same direction. The SOC
$\alpha\bk\cdot\boldsymbol{\sigma}$ thus reduces energy when the
spin $<\boldsymbol{\sigma}>$ is antiparallel with $\mathbf{j}$. This
average spin provide a spin transfer torque on the local
magnetizations, which is therefore analogous to a effective magnetic
field along $-\mathbf{j}$.

Although the effective field along the current will not change
routine observables such as the topological Hall
effect\cite{neubauer_topological_2009}, a physical consequence of
this inverse spin-galvanic effect is the current induced helix
reorientations in the helimagnet. It has been shown that the
hysteresis under low field is vanishingly small in B20
compounds\cite{bauer_magnetic_2012}. The orientation of the spin
helix is completely determined by the direction of external magnetic
field. Here we propose that one can use an electric current, instead
of magnetic fields, to orient the spin helix. The helix would
propagate in parallel with the current, which can be experimentally
confirmed by neutron scattering. This effect also applies in B20
thin film, where the Lorentz TEM would be a proper way to detect.

\section{Helical Resistance}

The SOC has various consequences in the magnetoresistances. Under a
low external magnetic field, the ground state of B20 is the helical
state, assembling a successive arrays of magnetic domains. The
domain wall resistance originated from collective spin scattering
has been extensively studied both experimentally and theoretically
in the context of conventional ferromagnets. SOC would apparently
alter the the spin scattering, and leads to unconventional helical
resistance in B20 compounds.

We consider the Hamiltonian in Eq. (\ref{fullHamiltonian}) while
including of the Zeeman term $\boldsymbol{\sigma} \cdot \mathbf{
h}$.
\begin{eqnarray}
H & = & \frac{k^2}{2m} - \mu + \mathbf{h} \cdot \boldsymbol{\sigma} + \alpha \mathbf{ k} \cdot \boldsymbol{\sigma} + \beta \Big( k_x \left( k_y^2 - k_z^2 \right) \sigma_x \nonumber  \\
& & +  k_y \left( k_z^2 - k_x^2 \right) \sigma_y +  k_z \left( k_x^2 - k_y^2 \right) \sigma_z   \Big) \label{Eqn:ZeemanHam} \\
U(\mathbf{ r}) & = & \frac1N \sum_i ( v\ \mathrm{I}_{2\times 2} - j
\mathbf{h} \cdot \boldsymbol{\sigma}) \delta_{\fvec r, \fvec R_i}
\label{Eqn:Imp}
\end{eqnarray}
Here, $\mathbf{ h}$ is the local magnetization. It rotates in $yz$
plane perpendicular to $x$, the propagation direction. $U$ is the
impurity potential, including both the scalar potential $v$ and
spin-dependent potential $-j \mathbf{h} \cdot
\boldsymbol{\sigma}$\cite{levy_resistivity_1997}. In the domain
wall, the direction of magnetization is nonuniform. Under the
assumption of slowly varying spin configuration (the helix period $d
\gg 1/k_f$), one can perform an $SU(2)$ gauge transformation $R$ to
the Hamiltonian so that the spin in the domain wall points along
$\hat e_z$ direction. $R$ is set to be
\begin{equation}
R=\exp (-i\theta (x)\sigma _{x}) \label{Eqn:GaugeSU2}
\end{equation}%
with $\theta(x) = 2\pi x/d$ where $d$ is the helix period. By this
rotation,
\begin{align*}
R^{-1}(\sigma \cdot \hat{h})R & =  \sigma _{z} \\
R^{-1}\frac{\hbar ^{2}\nabla ^{2}}{2m}R & = \frac{\hbar ^{2}\nabla ^{2}}{2m}-i%
\frac{\hbar ^{2}}{2m}\sigma _{x}(\partial _{x}\theta )\partial _{x}\\
&=\frac{\hbar ^{2}\nabla ^{2}}{2m}-i\frac{\pi \hbar ^{2}}{md}\sigma
_{x}\partial _{x} \\
R^{-1}\sigma _{x}\partial _{x}R & = -i\frac{\pi }{d}+\sigma_{x}\partial _{x} \\
R^{-1}\sigma _{y}\partial _{y}R & = (\cos \theta \sigma_{y}-\sin\theta\sigma _{z})\partial _{y} \\
R^{-1}\sigma _{z}\partial _{z}R &=(\cos \theta \sigma
_{z}+\sin\theta \sigma _{y})\partial _{z}
\end{align*}
The cubic SOC terms are transformed in a more complicated way.
However, under the large helix period approximation, $d \gg 1/k_f$
with $k_f$ the Fermi wavevector, it can be simplified as
\begin{align*}
R^{-1} \sigma_x k_x (k_y^2 - k_z^2) R  \approx & \sigma_x k_x (k_y^2 - k_z^2) \\
R^{-1} \sigma_y k_y (k_z^2 - k_x^2) R  \approx & \big( \sigma_y \cos\theta - \sigma_z \sin\theta  \big) k_y (k_z^2 - k_x^2) \\
R^{-1} \sigma_y k_z (k_x^2 - k_y^2) R  \approx & \big( \sigma_z
\cos\theta + \sigma_y \sin\theta \big) k_z (k_x^2 - k_y^2)
\end{align*}


To calculate the helical conductivity, we solve the Boltzmann
equation by a perturbation method. The deviation of electron
distribution function from the equilibrium one $f_1 = f - f_0$ is
expanded in terms of the spherical harmonic functions $Y_l^m(\theta,
\phi)$ up to $l = 5$. The details of the calculation can be found in
the appendix.

Fig~\ref{Fig:HelicalR} shows the ratio between the longitudinal
resistivities with current perpendicular to the domain wall (CPW)
and current in the domain wall (CIW). It is found that the ratio is
strongly suppressed by the presence of the SOC. In addition, the
presence of cubic SOC terms only quantitatively change the ratio.

\begin{figure}
\includegraphics[width=2.5in]{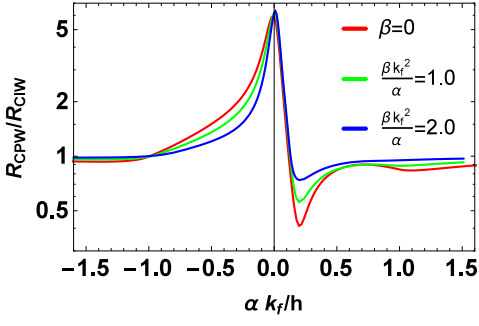}
\caption{The resistance ratio $R_{CPW}/R_{CIW}$ vs the spin orbital
coupling. In the plot, $\pi k_f/(m d h) = 0.2$, $j h/v = 1.06$.
Different curves are for different cubic SOC terms. It is clear that
the ratio approaches to the one when the SOC is much larger than the
Zeeman energy $h$.} \label{Fig:HelicalR}
\end{figure}
The minimum of the ratio is reached when \begin{equation} \alpha =
\frac{\pi}{m d} \ . \end{equation} To understand this minimum, we
consider the Hamiltonian with the linear SOC coupling only, ie.
$\beta = 0$. After the local $SU(2)$ gauge transformation, the
Hamiltonian becomes
\begin{align}
  H  = & H_0 + V + \mathrm{const.} \label{Eqn:Hamiltonian} \\
  H_0  = & \frac{p^2}{2m} - h \sigma_z \nonumber \\
  V  = & \left( \alpha - \frac {\pi}{m d} \right) \sigma_x p_x + \alpha \sigma_y \left( p_y \cos\theta + p_z \sin\theta \right) \nonumber \\
  & + \alpha \sigma_z \left( p_z \cos\theta - p_y \sin\theta  \right) \\
  \mathrm{const} = &  \frac{\pi^2}{8m d^2} - \frac{\alpha \pi}{2 d} \nonumber \ .
\end{align}
To simplify the notation, we define $\gamma = \alpha - \frac {\pi}{m
d}$. If SOC terms $\alpha k_f$, $\beta k_f^3$ are much smaller than
the Zeeman energy, the eigenstate of the Hamiltonian can be solved
based on perturbation. In addition, we assume that the impurity
scattering is strongly spin-dependent, ie.\ $v \approx j h$. In this
case, the conductivity is dominated by the outer Fermi surface, with
its intraband impurity scattering given by
\begin{equation}
  \left| M_{k\rightarrow k'}^{++} \right|^2 \approx  (v+j h)^2 \left[ \frac{\gamma^2 k_x k_x' + \frac{\alpha^2}2 (k_y k_y' + k_z k_z')}{4 h^2} \right]^2
\end{equation}
Here, $+$ is for outer Fermi surface. When $\alpha$ vanishes, it is
clear that scattering is larger at larger $k_x$ and $k_x'$. The
resistivity comes from the fermions with larger momentum along the
direction of the current. Therefore, $\rho_x \gg \rho_{y/z}$, and
thus the ratio $R_{CPW}/R_{CIW}$ reaches its maximum. When $\gamma =
0$ ($\alpha = \pi/(m d)$), the scattering rate is larger for larger
$k_{y/z}$ and $k'_{y/z}$. Therefore, $\rho_{y/z} \gg \rho_x$, and
the ratio $R_{CPW}/R_{CIW}$ reaches the minimum. More detailed
information can be found in the appendix.

\section{Anomalous Anisotropic Magnetoresistance}

In previous sections, we have mainly focused on the effects of the
linear SOC in Eq. (\ref{fullHamiltonian}). The cubic SOC only
slightly modifies these effects. The cubic term fails to bring any
qualitative change in these experiments. However from the symmetry's
point of view, only the cubic SOC breaks the $C_4$ rotation, and it
must give rise to anisotropic behaviors of the magnetoresistance. To
this end, we study the case when external magnetic field is
sufficiently large to polarize all magnetic moments along its
direction, and calculate the magnetoresistances. It is well known
that for most of the cubic ferromagnets such as Ni, the anisotropic
magnetoresistance (AMR) shows four fold symmetry when the magnetic
field rotates in the plane perpendicular to the current. However it
has already been reported that in B20 compounds such as
Fe$_{1-x}$Co$_x$Si\cite{huang_magnetoresistance_2014}, the AMR shows
anomalous behavior that only a two fold symmetry is respected. In
this section, we will show how the cubic SOC leads to this
observation.

The same model as the previous section is employed. The Hamiltonian
and the impurity potential is in the same form as Eq.
(\ref{Eqn:ZeemanHam}) and Eq. (\ref{Eqn:Imp}). The conductivity
$\sigma_{zz}(\mathbf{h})$ is calculated while varying the
magnetization $\mathbf{h}$ in the x-y plane. Most of the
ferromagnetic materials have high symmetry and respect $C_4$
symmetry. Thus $\sigma_{zz}(h \hat e_x)=\sigma_{zz}(h \hat e_y)$ in
these cases. However due to the $C_4$ breaking in B20 compounds, the
anomalous AMR is expected where the ratio between two conductivities
$\sigma_{zz}(h\hat{e}_x)$ and $\sigma_{zz}(h\hat{e}_y)$ deviates
from $1$.

Before calculation, let's explore the symmetries of the Hamiltonian
in Eq. (\ref{Eqn:ZeemanHam}). It is found that \be H(\alpha, \beta,
h \hat e_y) = S H(\alpha, -\beta, h \hat e_x) S^{-1} \ee Here $S$ is
the operator which rotates the system along $z$ axis by $\pi/2$.
Therefore, \be \sigma_{zz}(\alpha, \beta, h \hat e_x) =
\sigma_{zz}(\alpha, -\beta, h \hat e_y) \label{Eqn:AMRSym1} \ee In
addition, the conductivity is invariant under the space inversion
symmetry $P$, \be
 & & H(-\alpha, -\beta, \fvec h) = P H(\alpha, \beta, \fvec h) P^{-1} \nonumber  \\
  & \Longrightarrow & \sigma_{zz}(\alpha, \beta, \fvec h) = \sigma_{zz}(-\alpha, -\beta, \fvec h)
\ee Combined with Eq. (\ref{Eqn:AMRSym1}), it is concluded that \be
\sigma_{zz}(\alpha, \beta, h \hat e_x) = \sigma_{zz}(\alpha, -\beta,
h \hat e_y) = \sigma_{zz}(-\alpha, \beta, h \hat e_y)
\label{Eqn:AMRSym} \ee Solely by symmetry argument, it is found that
the anomalous AMR, defined as $\sigma_{zz}(h \hat e_x)/\sigma_{zz}(h
\hat e_y) - 1$, vanishes if either of two spin orbital couplings,
$\alpha$ and $\beta$, vanishes. Similarly, we have
$\sigma_{zz}(\alpha, \beta, h \hat e_h) = \sigma_{zz}(\alpha, \beta,
-h \hat e_h)$.

In this section, the conductivity is calculated in the same way as
the previous section. The Boltzmann equation is solved by
perturbation method. The deviation of the electron distribution
function is expanded by spherical harmonic functions up to $l = 5$.
The anisotropy comes from two different sources. (i) the Fermi
surface is anisotropic since the Hamiltonian in Eq.
(\ref{Eqn:ZeemanHam}) breaks $C_4$ symmetry. (ii) the eigenstate on
the Fermi surface is anisotropic, and thus leads to the anisotropic
impurity scattering by the spin-dependent potential in Eq.
(\ref{Eqn:Imp}).

The energy of the Hamiltonian in Eq. (\ref{Eqn:ZeemanHam}) is given
by
\begin{eqnarray}
\varepsilon _{F} & = & \frac{\hbar ^{2} k_{F}^{2}}{2m} \pm \Big( h_{z}^{2} + \alpha^{2}(k_{x}^{2}+k_{y}^{2}+k_{z}^{2}) \nonumber \\
& & + \beta
^{2}[k_{x}^{2}(k_{y}^{2}-k_{z}^{2})^{2}+k_{y}^{2}(k_{z}^{2}-k_{x}^{2})^{2}+k_{z}^{2}(k_{x}^{2}-k_{y}^{2})^{2}]  \nonumber \\
& & +  2h_{z}k_{z}[\alpha +\beta (k_{x}^{2}-k_{y}^{2})] \Big)^{1/2}
\label{Eqn:Energy}
\end{eqnarray}
The system contains two Fermi surfaces (FSs) since the Kramers
degeneracy is lifted. All terms in Eq. (\ref{Eqn:Energy}) respect
the symmetry exchanging the indices $x$ and $y$ except the last term
in square root. It shows explicitly that the Fermi surface asymmetry
is possible only when then magnetization is nonzero.

\begin{figure}[htbp]
\centering
\subfigure[\label{Fig:Scat:Asym}]{\includegraphics[width=1.6in]{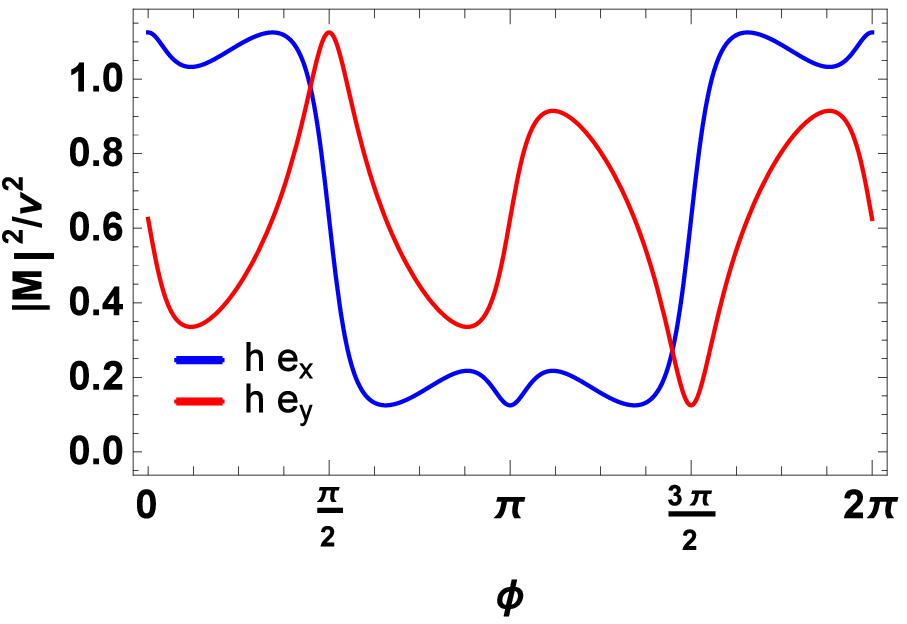}}
\subfigure[\label{Fig:Scat:Sym}]{\includegraphics[width=1.6in]{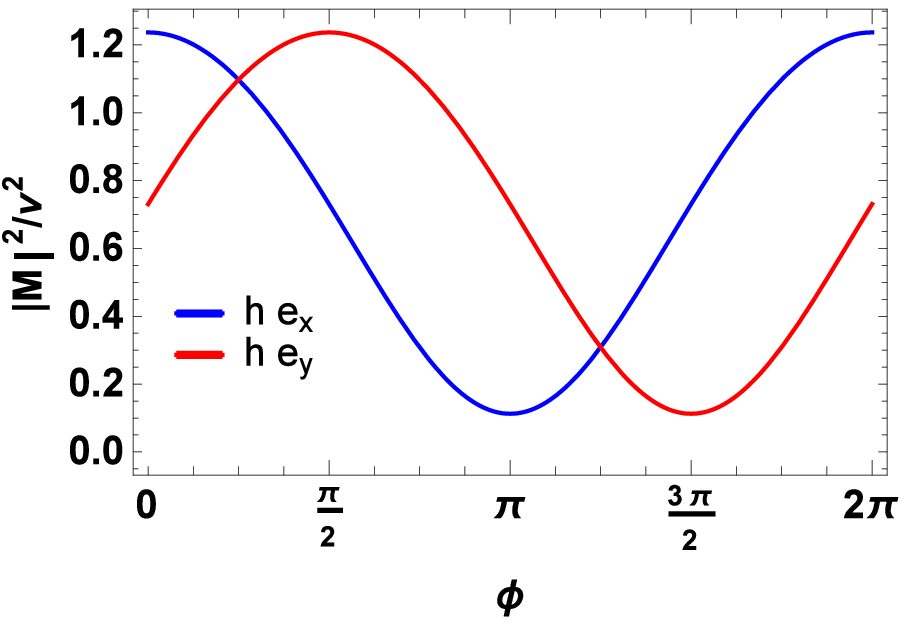}}
\caption{The intraband impurity scattering amplitude from $(0, 0,
k_f)$ to $(k_f \cos\phi, k_f\sin\phi, 0)$ on the outer Fermi
surface. Red and blue curves are the scattering magnitude as a
function of $\phi$ when the magnetization field points along $\hat
y$ and $\hat x$ respectively, respectively. (a) The Zeeman energy is
artificially turned off, while the impurity spin dependent is kept
to be nonzero with  $v/(j h) = 2.0$ and $\beta k_f^2/\alpha = 1.5$.
The $C_4$ symmetry is broken in the impurity scattering amplitude.
(b) $\beta=0$. Both the Zeeman energy and the impurity
spin-dependent potential are nonzero with $h/(\alpha k_f) = 0.1$,
and $v/(j h) = 2.0$. It is clear that $C_4$ symmetry recovers for
$\beta = 0$. } \label{Fig:Scat}
\end{figure}
However, the impurity scattering asymmetry is present as long as
$\alpha$ and $\beta$ are non-zero. Fig.~\ref{Fig:Scat} shows the
intraband scattering magnitude $|M|^2$ from $(0, 0, k_f)$ to $(k_f
\cos\phi, k_f \sin\phi, 0)$ on the outer Fermi surface. Red and blue
curves are $|M|^2$ as a function of $\phi$ when the impurity
magnetization points along $\hat y$ and $\hat x$, respectively. If
the $C_4$ symmetry is kept, the blue curve should be the same as the
red one after a translation of $\pi/2$, which is apparently not the
truth. It is noticeable that even when Zeeman field vanishes, the
impurity scattering still breaks the $C_4$ symmetry, although the
shape of Fermi surface is still $C_4$ symmetric. Of course in
reality, both the Zeeman term and spin-dependant scattering coexist.

\begin{figure}[htbp]
\centering \vspace{0.5cm}
\subfigure[\label{Fig:AMR:SOC}]{\includegraphics[width=1.6in]{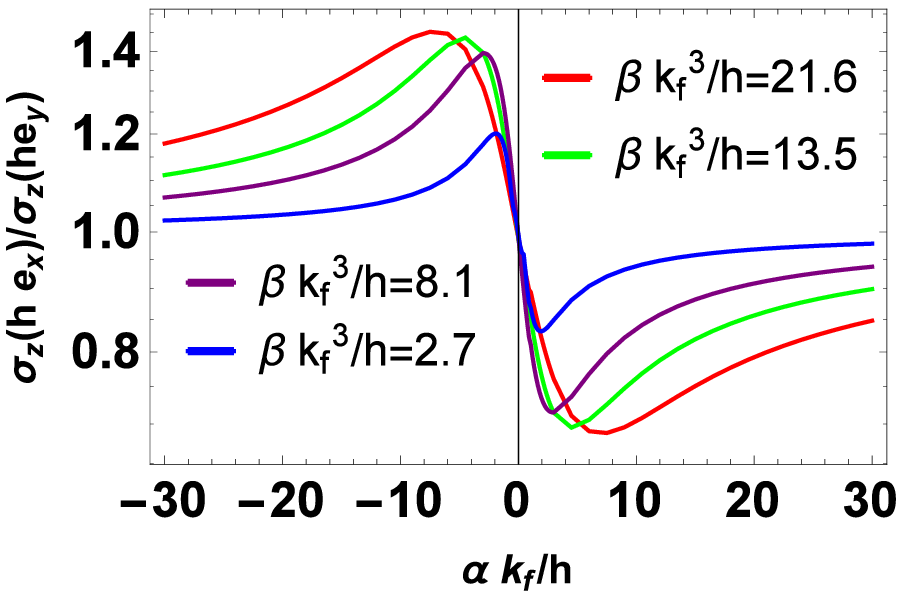}}
\subfigure[\label{Fig:AMR:H}]{\includegraphics[width=1.6in]{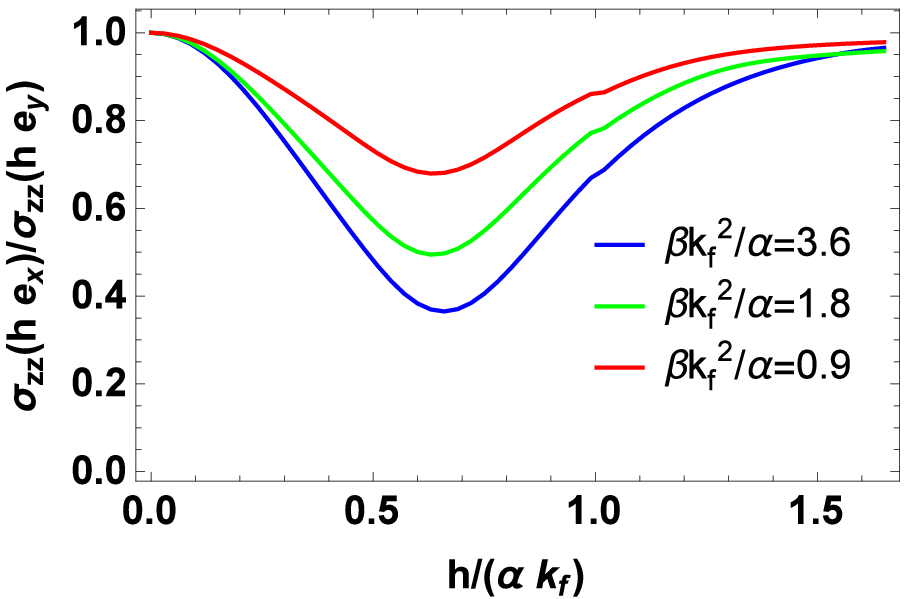}}
\caption{The ratio $\sigma_z(h \hat e_x)/\sigma_z(h \hat e_y)$ vs
SOC couplings and the magnetization field. (a) The ratio as a
function of SOC with $v/(j h) = 2.0$. It becomes one when $\alpha =
0$ or $\beta k_f^2/\alpha \rightarrow 0$. This agrees with our
conclusion in Eq.(\ref{Eqn:AMRSym}) by symmetry argument. (b) The
ratio vs Zeeman energy $h$ with $j=1.0$. The anomalous AMR vanishes
when $h=0$ or $h \rightarrow \infty$.} \label{Fig:AMR}
\end{figure}


Fig~\ref{Fig:AMR:SOC} shows the ratio $\sigma_{zz}(h \hat
e_x)/\sigma_{zz}(h \hat e_y)$ as a function of the SOC. The ratio is
smaller than $1$ when $\alpha$ and $\beta$ have the same sign, and
larger than $1$ when two SOCs have different signs. This is
consistent with the conclusion Eq. (\ref{Eqn:AMRSym}) derived by
symmetry arguments. In addition, it is found that the anomalous AMR
vanishes when either $\alpha = 0$ or $\alpha \rightarrow \infty$.
The latter implies that the anomalous AMR vanishes when $\beta
\rightarrow 0$. This result agrees with our physical picture based
on the Fermi surface topology. $C_4$ symmetry is restored on the
Fermi surface when $\beta \rightarrow 0$.

Fig~\ref{Fig:AMR:H} shows the anomalous AMR as a function of the
magnetization field $h$. When $h$ vanishes, not only the Zeeman
energy vanishes, but also the impurity potential becomes
spin-independent. Therefore, the impurity scattering becomes $C_4$
symmetric, as well as the Fermi surface. In this case, anomalous AMR
vanishes. Our calculations agree well with the experimental
results\cite{huang_magnetoresistance_2014}. When the temperature is
raised above the Curie temperature, the anomalous AMR vanishes. This
corresponds to the case with vanishing magnetization $\mathbf{h}$.
In another limit when $h \rightarrow\infty$, the spin on the Fermi
surface is fixed by the Zeeman energy. In this situation, the
impurity scattering and Fermi surface become isotropic, and
therefore the anomalous AMR vanishes.

\begin{figure}[htbp]
\centering \vspace{0.5cm}
\includegraphics[width=3.2in]{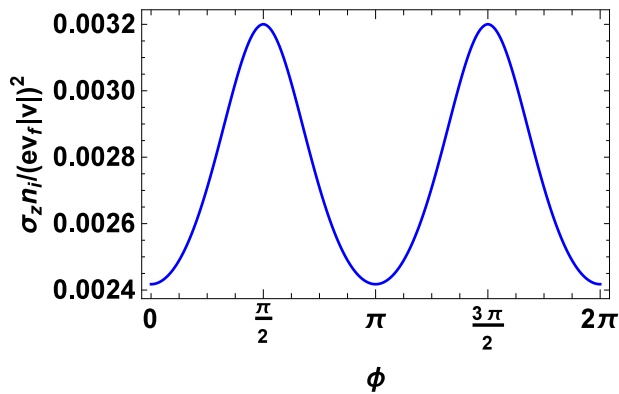}
\caption{Conductivity vs direction of in-plane magnetization field.
In the plot, $\alpha k_f/h = 3.0$, $\beta k_f^3/h = 5.4$, and $v/( j
h) = 2.0$. $n_i$ is the impurity density.} \label{Fig:CondPhi}
\end{figure}

Fig~\ref{Fig:CondPhi} shows the variation of the conductivity
$\sigma_{zz}$ as the direction of in-plane magnetization field
changes. Note that $\sigma_{zz}$ reaches its minimum(maximum) when
the field points along $x$($y$) direction. In our calculation, this
comes from that assumption that two SOC couplings have the same
sign. If $\alpha$ and $\beta$ have different signs, the
minimum(maximum) is reached when $\boldsymbol{h}$ is along $y$($x$)
axis. This result reproduces the experimental observations in
\cite{huang_magnetoresistance_2014}.

\section{conclusion}

In conclusion, the spin-orbital coupling is the fountain of various
interesting phenomena in B20 compounds. It not only provides the
antisymmetric spin interactions, but also dramatically change
behaviors of the electron transport. The effective Hamiltonian
constructed in this work captures the main effects of the SOC in
conduction electrons. Despite its simple form, the emergent new
physics is closely associated with several experiments. It also
calls for bunch of future works to study the spin transports related
to this Hamiltonian. First principle calculations are also
encouraged to determine the strength of the SOCs.

We are grateful for fruitful discussions with C. L. Chien, N.
Nagaosa, S. X. Huang, C. X. Liu, O. Tchernyshyov, R. M. Fernandes,
V. Galitski, and R. B. Tao. JK was supported by the Office of Basic
Energy Sciences U.S. Department of Energy under awards numbers
DE-SC0012336. JZ was supported by the TIPAC, by the U.S. Department
of Energy under Award DEFG02-08ER46544, and by the National Science
Foundation under Grant No. ECCS-1408168.

\newpage
\widetext \vspace{0.5cm}
\begin{center}
\textbf{\large Supplementary for ``Transport Theory of Metallic B20
Helimagnets''}
\end{center}
\setcounter{equation}{0} \setcounter{figure}{0}
\setcounter{table}{0} \setcounter{section}{0}
\makeatletter
\renewcommand{\theequation}{S\arabic{equation}}
\renewcommand{\thefigure}{S\arabic{figure}}

\section{Solving Boltzmann Equation by Perturbation}
If the Hamiltonian of the electron depends on the spin, the Kramer
degeneracy will be lifted in the presence of a magnetic field. Even
the single orbital Hamiltonian contains two different bands. The
impurity potential induce not only intraband scattering, but also
interband scattering. These scattering are, in general, highly
anisotropic and therefore lead to many interesting phenomena.
Although the relaxation time approximation, based on the assumption
of isotropic system, may still be able to explain experiments
qualitatively\cite{DomainWall}, it is questionable to produce a
reliably quantitative description. We have to solve the Boltzmann
equation without making any other assumptions. This section will
describe a perturbative way to solve this equation. Note that the
method presented here is not new, and has already been well
explained in the textbook\cite{Ziman}.

\subsection{Theory}
We assume $f(k) = f_0(k) + f_1(k)$, where $f_0$ is the Fermi
distribution function, or the distribution function at equilibrium.
The Boltzmann equation can be written as
\begin{align}
e \fvec E \cdot \fvec v_+ (k) \left( \frac{\partial f_0}{\partial \epsilon} \right)_{\epsilon = \epsilon_+(k)} & = 2\pi n_i \int \frac{\rmd^3 k'}{(2\pi)^3} \Big\{ \left( f_1^+(k') -f_1^+(k) \right) T^{++}(k, k') \delta\big(\epsilon_+(k) - \epsilon_+(k') \big) + \nonumber \\
&  \left( f_1^-(k') -f_1^+(k) \right) T^{+-}(k, k') \delta\big(\epsilon_+(k) - \epsilon_-(k') \big) \Big\} \\
e \fvec E \cdot \fvec v_- (k) \left( \frac{\partial f_0}{\partial \epsilon} \right)_{\epsilon = \epsilon_-(k)} & = 2\pi n_i \int \frac{\rmd^3 k'}{(2\pi)^3} \Big\{ \left( f_1^+(k') -f_1^-(k) \right) T^{-+}(k, k') \delta\big(\epsilon_-(k) - \epsilon_+(k') \big) + \nonumber \\
&  \left( f_1^-(k') -f_1^-(k) \right) T^{--}(k, k')
\delta\big(\epsilon_-(k) - \epsilon_-(k') \big) \Big\} \ ,
\end{align}
where the super(sub)scripts ``$+/-$" refers to the two different
bands due to removing Kramer degeneracy in the single electron
Hamiltonian. $T^{\sigma,\sigma'}(k, k')$ is the scattering matrix
element for electron from the state of $|\sigma, k\rangle$ to the
state of $|\sigma', k' \rangle$. Thus, it contains both the
intra-band and interband scattering. It is very easy to generalize
to the Hamiltonian with more than two bands. Here, we will try the
solution in the form of
\[ f^{\pm}_1(k) = g_{\pm}(k) \left( - \frac{\partial f_0}{\partial \epsilon} \right)_{\epsilon = \epsilon_{\pm}(k)} \ . \]
The Boltzmann equation becomes
\begin{align}
- e \fvec E \cdot \fvec v_+ (k)  & = \frac{n_i}{(2\pi)^2} \int \rmd^3 k' \Big\{ \left( g_+(k') - g_+(k) \right) T^{++}(k, k') \delta\big(\epsilon_+(k) - \epsilon_+(k') \big) \nonumber \\
&  \left( g_-(k') - g_+(k) \right) T^{+-}(k, k') \delta\big(\epsilon_+(k) - \epsilon_-(k') \big) \Big\} \\
- e \fvec E \cdot \fvec v_- (k) & =  \frac{n_i}{(2\pi)^2} \int \rmd^3 k' \Big\{ \left( g_+(k') - g_-(k) \right) T^{-+}(k, k') \delta\big(\epsilon_-(k) - \epsilon_+(k') \big) \nonumber \\
&  \left( g_-(k') - g_-(k) \right) T^{--}(k, k') \delta\big(\epsilon_-(k) - \epsilon_-(k') \big) \Big\} \ \\
\fvec j & = e \int \frac{\rmd^3 k}{(2\pi)^3} \left( \fvec v_+(k)
g_+(k) \left( - \frac{\partial f_0}{\partial \epsilon}
\right)_{\epsilon = \epsilon_{+}(k)} + \fvec v_-(k) g_-(k) \left( -
\frac{\partial f_0}{\partial \epsilon} \right)_{\epsilon =
\epsilon_-(k)} \right) \ ,
\end{align}
where $n_i$ is the density of impurity. To further simplify the
notation, we can define the state vector and the linear operator
\begin{align}
  \Phi(k) & = \begin{pmatrix} g_+(k) \\ g_-(k)  \end{pmatrix}   \\
  X(k) & = - e \begin{pmatrix} \fvec E \cdot \fvec v_+(k) \\ \fvec E \cdot \fvec v_-(k)  \end{pmatrix} \\
  (W \Phi)_+(k) & = \frac{n_i}{(2\pi)^2} \int\rmd^3 k' \ \Big\{ \left( g_+(k') - g_+(k) \right) T^{++}(k, k') \delta\big(\epsilon_+(k) - \epsilon_+(k') \big) + \nonumber \\
  & \hspace{1cm} \left( g_-(k') - g_+(k) \right) T^{+-}(k, k') \delta\big(\epsilon_+(k) - \epsilon_-(k') \big) \Big\} \label{Eqn:WPhiP} \\
  (W \Phi)_-(k) & = \frac{n_i}{(2\pi)^2} \int\rmd^3 k' \ \Big\{ \left( g_+(k') - g_-(k) \right) T^{-+}(k, k') \delta\big(\epsilon_-(k) - \epsilon_+(k') \big) + \nonumber \\
  & \hspace{1cm} \left( g_-(k') - g_-(k) \right) T^{--}(k, k') \delta\big(\epsilon_-(k) - \epsilon_-(k') \big) \Big\} \label{Eqn:WPhiM}
\end{align}
It is clear that the Boltzmann equation can be written as \be
 W \Phi(k) = X(k) \ . \label{Eqn:BZ}
\ee

\subsection{Approximation}
We will work at $T = 0$, so that the derivative of Fermi
distribution function becomes a delta function. We will integrate
our the magnitude of momentum and get the term $k^2/v_k$. Therefore,
it is more convenient to define the ``modified" scattering as
\[ S^{\sigma \sigma'}_{k k'} =  T^{\sigma \sigma'}_{k k'} \left( \frac{k^2}{v_k} \right)_{\sigma} \left( \frac{k^2}{v_k} \right)_{\sigma'} \qquad  Z_{\hat n}(k) = \begin{pmatrix} v^+_n \left( \frac{k^2}{v_k} \right)_+ \\ v^-_n \left( \frac{k^2}{v_k} \right)_-  \end{pmatrix}\ .  \]
Clearly, in this definition, $S$ is symmetric, ie.\ $S^{\sigma
\sigma'}_{k k'} = S^{\sigma' \sigma}_{k' k}$. In practice, we will
expand the wave vector $\Phi(k)$ in terms of spherical harmonic
functions. Define the matrix
\[ M(k) = \begin{pmatrix}
  \left( \dfrac{k^2}{v_k} \right)_+ & 0 \\ 0 & \left( \dfrac{k^2}{v_k} \right)_-
\end{pmatrix} \ ,  \]
and multiply it on both sides of Boltzmann equation. \be
  & & \sum_{\sigma' l' m'}\left(  S^{\sigma \sigma'}(l,m; l',m') + (-)^{m+1} \delta_{\sigma \sigma'} \sum_{l_1} \Big( \sum_{\sigma_1} S^{\sigma \sigma_1}(l_1,m-m';0,0) \Big) B(l_1,m-m';l',m';l)  \right) g_{\sigma'}(l' m')  \nonumber \\
   & = & Z_{\hat n}^{\sigma}(l, m)   \label{Eqn:BZMatrix}
\ee
\begin{align*}
  g_{\sigma}(\hat k) & = \sum_{l m} g_{\sigma}(l, m) Y_l^m(\hat k) \ , \qquad Z_{\hat n}^{\sigma}(\hat k)  = \sum_{l m} Z_{\hat n}^{\sigma}(l,m) Y_l^m(\hat k) \\
  S^{\sigma \sigma'}(\hat k, \hat k') & = \sum_{l m} \sum_{l' m'} S^{\sigma \sigma'}(l,m;l',m')\ Y_l^m(\hat k) \left( Y_{l'}^{m'}(\hat k') \right)^* \\
  B(l_1,m-m';l',m';l) & = (-)^m \sqrt{(2l_1 + 1)(2l' +1)(2l+1)} \begin{pmatrix} l_1 & l' & l \\m - m' & m' & -m  \end{pmatrix} \begin{pmatrix} l_1 & l' & l \\0 & 0 & 0  \end{pmatrix} \\
  \sigma_n & = \frac1{(2\pi)^3} \sum_{\sigma l m} (-)^m Z_{\hat n}^{\sigma}(l,m) g_{\sigma}(l, -m)
\end{align*}
Note that the Wigner $3j$ symbol is used. Now, it is clear that we
can solve the Boltzmann equations in an approximate way.

In practice, we will set a cutoff $l_c$ on the angular momentum of
the spherical harmonic functions. Larger $l_c$ produces more precise
solutions. In our calculation, we set $l_c=5$ for limitations on the
computing source. We will see that it produces quantitatively
different results from the relaxation time approximation.

\section{Helical Resistance}
In this section, we will study the resistance of helical phase in
the presence of the spin-orbital coupling. Apply the local $SU(2)$
gauge transformation in Eqn~\ref{Eqn:GaugeSU2}, the cubic SOC term
becomes
\begin{align*}
  R^{-1} \sigma_x k_x (k_y^2 - k_z^2) R  = & \sigma_x \left( k_x - \frac{2\pi}d \sigma_x \right)(k_y^2 - k_z^2) \\
R^{-1} \sigma_y k_y (k_z^2 - k_x^2) R  \approx & \big( \sigma_y \cos\theta - \sigma_z \sin\theta  \big) k_y \left( k_z^2 - \left( k_x - \frac{2\pi}d \sigma_x \right)^2 \right) \\
R^{-1} \sigma_y k_z (k_x^2 - k_y^2) R  \approx & \big( \sigma_z
\cos\theta + \sigma_y \sin\theta \big) k_z \left( \left( k_x -
\frac{2\pi}d \sigma_x \right)^2 - k_y^2 \right)
\end{align*}
With the assumption of $1/d \ll k_f$, it is safe to ignore the extra
$2\pi/d$ terms in the transformed cubic SOC.

Fig~\ref{Fig:HelicalR} shows the ratio $R_{CPW}/R_{CIW}$ as a
function of SOC. It is found that this ratio peaks when SOC vanishes
and reaches its minimum when $\alpha = \pi h/(m d)$ or $\gamma = 0$.
It is argued that the peaks and troughs are related with the
anisotropy of the impurity scattering. In this section, we
investigate the scattering more systematically, and reveals how the
anisotropy of the scattering is changed by SOC.

As in the main text, we consider the case $\beta = 0$ for
simplicity. The Hamiltonian is
\begin{align*}
  H  = & H_0 + V  \\
  H_0  = & \frac{p^2}{2m} - h \sigma_z \\
  V  = & \gamma \sigma_x p_x + \alpha \sigma_y \left( p_y \cos\theta + p_z \sin\theta \right) + \alpha \sigma_z \left( p_z \cos\theta - p_y \sin\theta  \right) \\
  U = & v\ \mathrm{I}_{2\times 2} - j h \sigma_z
\end{align*}
Here $\gamma = \alpha - \pi/(m d)$ and $U$ is the transformed
impurity potential. For small SOC $\alpha k_f \ll h$, we treat $V$
as the perturbation and calculate the wavefunction.
\begin{align*}
  \vert +, k \rangle & = N_+^{-1} (k_+) \left[ \begin{pmatrix} e^{i k_+ r} \\ 0 \end{pmatrix} - \dfrac{\gamma k_x + i \alpha \left( k_y \cos\theta + k_z \sin\theta \right)}{2h} \begin{pmatrix} 0 \\ e^{i k_+ r}  \end{pmatrix} \right] \\
  \vert -, k \rangle & = N_-^{-1} (k_+) \left[ \begin{pmatrix} 0 \\ e^{i k_- r} \end{pmatrix} + \dfrac{\gamma k_x - i \alpha \left( k_y \cos\theta + k_z \sin\theta \right)}{2h} \begin{pmatrix} e^{i k_+ r} \\ 0 \end{pmatrix} \right]
\end{align*}
$N_{\pm}$ are introduced for normalization. In principle, $\theta =
2\pi x/d$ is a space dependent function, and thus will mix the plane
wave with different momentum and different fermi surfaces. Here, it
is assumed that the space varying is much smaller than the
difference of two fermi momentum. Especially, $1/d \ll \vert| k_- -
k_+ \vert$, ie.\ it will not mix two Fermi surfaces. Now, we
calculate the scattering matrix elements in the case of strongly
spin-dependent impurity potential, eg.\ $v \approx j h$. After
integrating over $\theta(x)$.
\begin{align}
  T^{++}(k, k')  & = \left| \langle + k'|U| + k \rangle \right|^2 \approx  \left( \frac{v +j h}{(2 h)^2} \right)^2 \left( \gamma^2 k_x k_x' + \frac{\alpha^2}2 \left[ k_y k_y' + k_z k_z' \right] \right)^2 \label{Eqn:PPScatter} \\
  T^{--}(k, k') & = \left| \langle - k'|U| - k \rangle \right|^2 \approx (v + j h)^2  \nonumber \\ 
  T^{+-}(k, k') & = \left| \langle - k'|U| + k \rangle \right|^2 \approx \frac{\gamma^2}{(2h)^2} \left[ (v+j h) k_x - (v-j h) k_x' \right]^2 + \frac12 \frac{\alpha^2}{(2h)^2} \left[ \left( (v + j h) k_y + (v - j h)k_y' \right)^2 + (k_y \leftrightarrow k_z) \right]   \nonumber  
\end{align}
It is easy to see that $T^{++} \ll T^{--}$. The conductivity is
dominated by the fermi surface with smaller intraband scattering as
those parts with large scattering rate will be ``shorted out"
\cite{BZShort}. Therefore, we focus only on $T^{++}$ here.

\begin{figure}[htbp]
\includegraphics[width=3.5in]{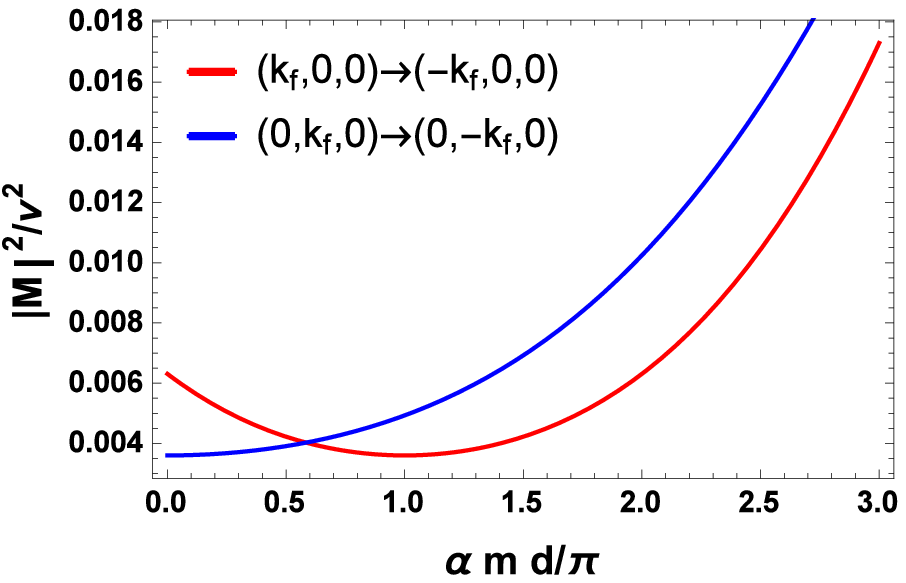}
\caption{Scattering magnitude vs.\ the spin orbital coupling
$\alpha$ with $j h/v= 1.06$, and $\beta = 0$. Red: Scattering
between $(k_f, 0, 0)$ and $(-k_f, 0, 0)$. It is clear that the
scattering rate reaches its minimum when $\alpha \approx \pi/(m d)$.
This behavior is typical for the scattering between large $k_x$,
which accounts for the resistivity when current is perpendicular to
the domain wall.  Blue: Scattering between $(0, k_f, 0)$ and $(0,
-k_f, 0)$. It monotonically increase with SOC. This is common among
the scattering between large $k_y$ ($k_z$) which accounts for the
resistivity when current is parallel to the domain wall. }
\label{Fig:ScatterDW}
\end{figure}

For currents perpendicular to the domain wall (along $\hat x$), the
electrons contributed to this flow are dominated by those with
larger $k_x$ since the velocity $v_x \approx k_x/m$. For currents
parallel to the domain wall, the electrons contributed to this flow
are dominated by those with larger $k_y$($k_z$) since the velocity
$v_x \approx k_x/m$ ($v_y \approx k_y/m$). When SOC vanishes $\alpha
= 0$, $\gamma \neq 0$, larger $k_x$ and $k_x'$ leads to larger
scattering in Eqn~\ref{Eqn:PPScatter}. Therefore, $R_{CPW} \gg
R_{CIW}$, and the ratio reaches its maximum. When $\alpha = \pi /(m
d)$, $\gamma = 0$, $\alpha \neq 0$, larger $k_y$ and $k_y'$ (or
$k_z$ and $k_z'$) leads to larger scattering. Therefore, $R_{CIW}
\gg R_{CPW}$, and the minimum of ratio emerges. When SOC becomes
very large, spin on the fermi surface is locked by the direction of
the fermi momentum. In this limit, the anisotropy of the scattering
vanishes and the ratio approaches to $1$, as shown in
Fig~\ref{Fig:ScatterDW}.


\begin{thebibliography}{38}
\expandafter\ifx\csname
natexlab\endcsname\relax\def\natexlab#1{#1}\fi
\expandafter\ifx\csname bibnamefont\endcsname\relax
  \def\bibnamefont#1{#1}\fi
\expandafter\ifx\csname bibfnamefont\endcsname\relax
  \def\bibfnamefont#1{#1}\fi
\expandafter\ifx\csname citenamefont\endcsname\relax
  \def\citenamefont#1{#1}\fi
\expandafter\ifx\csname url\endcsname\relax
  \def\url#1{\texttt{#1}}\fi
\expandafter\ifx\csname urlprefix\endcsname\relax\def\urlprefix{URL
}\fi \providecommand{\bibinfo}[2]{#2}
\providecommand{\eprint}[2][]{\url{#2}}

\bibitem[{\citenamefont{Schlesinger et~al.}(1993)\citenamefont{Schlesinger,
  Fisk, Zhang, Maple, DiTusa, and Aeppli}}]{schlesinger_unconventional_1993}
\bibinfo{author}{\bibfnamefont{Z.}~\bibnamefont{Schlesinger}},
  \bibinfo{author}{\bibfnamefont{Z.}~\bibnamefont{Fisk}},
  \bibinfo{author}{\bibfnamefont{H.-T.} \bibnamefont{Zhang}},
  \bibinfo{author}{\bibfnamefont{M.~B.} \bibnamefont{Maple}},
  \bibinfo{author}{\bibfnamefont{J.}~\bibnamefont{DiTusa}}, \bibnamefont{and}
  \bibinfo{author}{\bibfnamefont{G.}~\bibnamefont{Aeppli}},
  \bibinfo{journal}{Physical Review Letters} \textbf{\bibinfo{volume}{71}},
  \bibinfo{pages}{1748} (\bibinfo{year}{1993}),
  \urlprefix\url{http://link.aps.org/doi/10.1103/PhysRevLett.71.1748}.

\bibitem[{\citenamefont{DiTusa et~al.}(1997)\citenamefont{DiTusa, Friemelt,
  Bucher, Aeppli, and Ramirez}}]{ditusa_metal-insulator_1997}
\bibinfo{author}{\bibfnamefont{J.~F.} \bibnamefont{DiTusa}},
  \bibinfo{author}{\bibfnamefont{K.}~\bibnamefont{Friemelt}},
  \bibinfo{author}{\bibfnamefont{E.}~\bibnamefont{Bucher}},
  \bibinfo{author}{\bibfnamefont{G.}~\bibnamefont{Aeppli}}, \bibnamefont{and}
  \bibinfo{author}{\bibfnamefont{A.~P.} \bibnamefont{Ramirez}},
  \bibinfo{journal}{Physical Review Letters} \textbf{\bibinfo{volume}{78}},
  \bibinfo{pages}{2831} (\bibinfo{year}{1997}),
  \urlprefix\url{http://link.aps.org/doi/10.1103/PhysRevLett.78.2831}.

\bibitem[{\citenamefont{M{\"u}hlbauer et~al.}(2009)\citenamefont{M{\"u}hlbauer,
  Binz, Jonietz, Pfleiderer, Rosch, Neubauer, Georgii, and
  B{\"o}ni}}]{muhlbauer_skyrmion_2009}
\bibinfo{author}{\bibfnamefont{S.}~\bibnamefont{M{\"u}hlbauer}},
  \bibinfo{author}{\bibfnamefont{B.}~\bibnamefont{Binz}},
  \bibinfo{author}{\bibfnamefont{F.}~\bibnamefont{Jonietz}},
  \bibinfo{author}{\bibfnamefont{C.}~\bibnamefont{Pfleiderer}},
  \bibinfo{author}{\bibfnamefont{A.}~\bibnamefont{Rosch}},
  \bibinfo{author}{\bibfnamefont{A.}~\bibnamefont{Neubauer}},
  \bibinfo{author}{\bibfnamefont{R.}~\bibnamefont{Georgii}}, \bibnamefont{and}
  \bibinfo{author}{\bibfnamefont{P.}~\bibnamefont{B{\"o}ni}},
  \bibinfo{journal}{Science} \textbf{\bibinfo{volume}{323}},
  \bibinfo{pages}{915} (\bibinfo{year}{2009}), ISSN \bibinfo{issn}{0036-8075,
  1095-9203}, \urlprefix\url{http://www.sciencemag.org/content/323/5916/915}.

\bibitem[{\citenamefont{Pfleiderer et~al.}(2004)\citenamefont{Pfleiderer,
  Reznik, Pintschovius, L{\"o}hneysen, Garst, and
  Rosch}}]{pfleiderer_partial_2004}
\bibinfo{author}{\bibfnamefont{C.}~\bibnamefont{Pfleiderer}},
  \bibinfo{author}{\bibfnamefont{D.}~\bibnamefont{Reznik}},
  \bibinfo{author}{\bibfnamefont{L.}~\bibnamefont{Pintschovius}},
  \bibinfo{author}{\bibfnamefont{H.~v.} \bibnamefont{L{\"o}hneysen}},
  \bibinfo{author}{\bibfnamefont{M.}~\bibnamefont{Garst}}, \bibnamefont{and}
  \bibinfo{author}{\bibfnamefont{A.}~\bibnamefont{Rosch}},
  \bibinfo{journal}{Nature} \textbf{\bibinfo{volume}{427}},
  \bibinfo{pages}{227} (\bibinfo{year}{2004}), ISSN \bibinfo{issn}{0028-0836},
  \urlprefix\url{http://www.nature.com/nature/journal/v427/n6971/abs/nature022%
32.html}.

\bibitem[{\citenamefont{Ritz et~al.}(2013)\citenamefont{Ritz, Halder, Wagner,
  Franz, Bauer, and Pfleiderer}}]{ritz_formation_2013}
\bibinfo{author}{\bibfnamefont{R.}~\bibnamefont{Ritz}},
  \bibinfo{author}{\bibfnamefont{M.}~\bibnamefont{Halder}},
  \bibinfo{author}{\bibfnamefont{M.}~\bibnamefont{Wagner}},
  \bibinfo{author}{\bibfnamefont{C.}~\bibnamefont{Franz}},
  \bibinfo{author}{\bibfnamefont{A.}~\bibnamefont{Bauer}}, \bibnamefont{and}
  \bibinfo{author}{\bibfnamefont{C.}~\bibnamefont{Pfleiderer}},
  \bibinfo{journal}{Nature} \textbf{\bibinfo{volume}{497}},
  \bibinfo{pages}{231} (\bibinfo{year}{2013}), ISSN \bibinfo{issn}{0028-0836},
  \urlprefix\url{http://www.nature.com/nature/journal/v497/n7448/abs/nature120%
23.html}.

\bibitem[{\citenamefont{Huang and Chien}(2012)}]{huang_extended_2012}
\bibinfo{author}{\bibfnamefont{S.~X.} \bibnamefont{Huang}} \bibnamefont{and}
  \bibinfo{author}{\bibfnamefont{C.~L.} \bibnamefont{Chien}},
  \bibinfo{journal}{Phys. Rev. Lett.} \textbf{\bibinfo{volume}{108}},
  \bibinfo{pages}{267201} (\bibinfo{year}{2012}),
  \urlprefix\url{http://link.aps.org/doi/10.1103/PhysRevLett.108.267201}.

\bibitem[{\citenamefont{Yu et~al.}(2011)\citenamefont{Yu, anazawa, Onose,
  Kimoto, Zhang, Ishiwata, Matsui, and Tokura}}]{yu_near_2011}
\bibinfo{author}{\bibfnamefont{X.~Z.} \bibnamefont{Yu}},
  \bibinfo{author}{\bibfnamefont{N.~K.} \bibnamefont{anazawa}},
  \bibinfo{author}{\bibfnamefont{Y.}~\bibnamefont{Onose}},
  \bibinfo{author}{\bibfnamefont{K.}~\bibnamefont{Kimoto}},
  \bibinfo{author}{\bibfnamefont{W.~Z.} \bibnamefont{Zhang}},
  \bibinfo{author}{\bibfnamefont{S.}~\bibnamefont{Ishiwata}},
  \bibinfo{author}{\bibfnamefont{Y.}~\bibnamefont{Matsui}}, \bibnamefont{and}
  \bibinfo{author}{\bibfnamefont{Y.}~\bibnamefont{Tokura}},
  \bibinfo{journal}{Nat Mater} \textbf{\bibinfo{volume}{10}},
  \bibinfo{pages}{106 } (\bibinfo{year}{2011}),
  \urlprefix\url{http://www.nature.com/nmat/journal/v10/n2/abs/nmat2916.html}.

\bibitem[{\citenamefont{Seki et~al.}(2012)\citenamefont{Seki, Yu, Ishiwata, and
  Tokura}}]{seki_observation_2012}
\bibinfo{author}{\bibfnamefont{S.}~\bibnamefont{Seki}},
  \bibinfo{author}{\bibfnamefont{X.~Z.} \bibnamefont{Yu}},
  \bibinfo{author}{\bibfnamefont{S.}~\bibnamefont{Ishiwata}}, \bibnamefont{and}
  \bibinfo{author}{\bibfnamefont{Y.}~\bibnamefont{Tokura}},
  \bibinfo{journal}{Science} \textbf{\bibinfo{volume}{336}},
  \bibinfo{pages}{198 } (\bibinfo{year}{2012}),
  \urlprefix\url{http://www.sciencemag.org/content/336/6078/198.full.html}.

\bibitem[{\citenamefont{Uchida et~al.}(2006)\citenamefont{Uchida, Onose,
  Matsui, and Tokura}}]{uchida_real-space_2006}
\bibinfo{author}{\bibfnamefont{M.}~\bibnamefont{Uchida}},
  \bibinfo{author}{\bibfnamefont{Y.}~\bibnamefont{Onose}},
  \bibinfo{author}{\bibfnamefont{Y.}~\bibnamefont{Matsui}}, \bibnamefont{and}
  \bibinfo{author}{\bibfnamefont{Y.}~\bibnamefont{Tokura}},
  \bibinfo{journal}{Science} \textbf{\bibinfo{volume}{311}},
  \bibinfo{pages}{359} (\bibinfo{year}{2006}), ISSN \bibinfo{issn}{0036-8075,
  1095-9203}, \urlprefix\url{http://www.sciencemag.org/content/311/5759/359}.

\bibitem[{\citenamefont{Dzyaloshinsky}(1958)}]{dzyaloshinsky_thermodynamic_195%
8}
\bibinfo{author}{\bibfnamefont{I.}~\bibnamefont{Dzyaloshinsky}},
  \bibinfo{journal}{Journal of Physics and Chemistry of Solids}
  \textbf{\bibinfo{volume}{4}}, \bibinfo{pages}{241} (\bibinfo{year}{1958}),
  ISSN \bibinfo{issn}{0022-3697},
  \urlprefix\url{http://www.sciencedirect.com/science/article/pii/002236975890%
0763}.

\bibitem[{\citenamefont{Moriya}(1960)}]{moriya_anisotropic_1960}
\bibinfo{author}{\bibfnamefont{T.}~\bibnamefont{Moriya}},
  \bibinfo{journal}{Physical Review} \textbf{\bibinfo{volume}{120}},
  \bibinfo{pages}{91} (\bibinfo{year}{1960}),
  \urlprefix\url{http://link.aps.org/doi/10.1103/PhysRev.120.91}.

\bibitem[{\citenamefont{Skyrme}(1961)}]{skyrme_non-linear_1961}
\bibinfo{author}{\bibfnamefont{T.~H.~R.} \bibnamefont{Skyrme}},
  \bibinfo{journal}{Proceedings of the Royal Society of London. Series A.
  Mathematical and Physical Sciences} \textbf{\bibinfo{volume}{260}},
  \bibinfo{pages}{127} (\bibinfo{year}{1961}), ISSN \bibinfo{issn}{1364-5021,
  1471-2946},
  \urlprefix\url{http://rspa.royalsocietypublishing.org/content/260/1300/127}.

\bibitem[{\citenamefont{Yu et~al.}(2010)\citenamefont{Yu, Onose, Kanazawa,
  Park, Han, Matsui, Nagaosa, and Tokura}}]{yu_real-space_2010}
\bibinfo{author}{\bibfnamefont{X.~Z.} \bibnamefont{Yu}},
  \bibinfo{author}{\bibfnamefont{Y.}~\bibnamefont{Onose}},
  \bibinfo{author}{\bibfnamefont{N.}~\bibnamefont{Kanazawa}},
  \bibinfo{author}{\bibfnamefont{J.~H.} \bibnamefont{Park}},
  \bibinfo{author}{\bibfnamefont{J.~H.} \bibnamefont{Han}},
  \bibinfo{author}{\bibfnamefont{Y.}~\bibnamefont{Matsui}},
  \bibinfo{author}{\bibfnamefont{N.}~\bibnamefont{Nagaosa}}, \bibnamefont{and}
  \bibinfo{author}{\bibfnamefont{Y.}~\bibnamefont{Tokura}},
  \bibinfo{journal}{Nature} \textbf{\bibinfo{volume}{465}},
  \bibinfo{pages}{901} (\bibinfo{year}{2010}), ISSN \bibinfo{issn}{0028-0836},
  \urlprefix\url{http://www.nature.com/nature/journal/v465/n7300/full/nature09%
124.html}.

\bibitem[{\citenamefont{R{\"o}{\textbackslash}ssler
  et~al.}(2006)\citenamefont{R{\"o}{\textbackslash}ssler, Bogdanov, and
  Pfleiderer}}]{rossler_spontaneous_2006}
\bibinfo{author}{\bibfnamefont{U.~K.}
  \bibnamefont{R{\"o}{\textbackslash}ssler}},
  \bibinfo{author}{\bibfnamefont{A.~N.} \bibnamefont{Bogdanov}},
  \bibnamefont{and}
  \bibinfo{author}{\bibfnamefont{C.}~\bibnamefont{Pfleiderer}},
  \bibinfo{journal}{Nature} \textbf{\bibinfo{volume}{442}},
  \bibinfo{pages}{797} (\bibinfo{year}{2006}), ISSN \bibinfo{issn}{0028-0836},
  \urlprefix\url{http://www.nature.com/nature/journal/v442/n7104/full/nature05%
056.html}.

\bibitem[{\citenamefont{Kadowaki et~al.}(1982)\citenamefont{Kadowaki, Okuda,
  and Date}}]{kadowaki_magnetization_1982}
\bibinfo{author}{\bibfnamefont{K.}~\bibnamefont{Kadowaki}},
  \bibinfo{author}{\bibfnamefont{K.}~\bibnamefont{Okuda}}, \bibnamefont{and}
  \bibinfo{author}{\bibfnamefont{M.}~\bibnamefont{Date}},
  \bibinfo{journal}{Journal of the Physical Society of Japan}
  \textbf{\bibinfo{volume}{51}}, \bibinfo{pages}{2433} (\bibinfo{year}{1982}),
  ISSN \bibinfo{issn}{0031-9015},
  \urlprefix\url{http://journals.jps.jp/doi/abs/10.1143/JPSJ.51.2433}.

\bibitem[{\citenamefont{Du et~al.}(2014)\citenamefont{Du, DeGrave, Xue, Liang,
  Ning, Yang, Tian, Zhang, and Jin}}]{du_highly_2014}
\bibinfo{author}{\bibfnamefont{H.}~\bibnamefont{Du}},
  \bibinfo{author}{\bibfnamefont{J.~P.} \bibnamefont{DeGrave}},
  \bibinfo{author}{\bibfnamefont{F.}~\bibnamefont{Xue}},
  \bibinfo{author}{\bibfnamefont{D.}~\bibnamefont{Liang}},
  \bibinfo{author}{\bibfnamefont{W.}~\bibnamefont{Ning}},
  \bibinfo{author}{\bibfnamefont{J.}~\bibnamefont{Yang}},
  \bibinfo{author}{\bibfnamefont{M.}~\bibnamefont{Tian}},
  \bibinfo{author}{\bibfnamefont{Y.}~\bibnamefont{Zhang}}, \bibnamefont{and}
  \bibinfo{author}{\bibfnamefont{S.}~\bibnamefont{Jin}}, \bibinfo{journal}{Nano
  Letters} \textbf{\bibinfo{volume}{14}}, \bibinfo{pages}{2026}
  (\bibinfo{year}{2014}), ISSN \bibinfo{issn}{1530-6984},
  \urlprefix\url{http://dx.doi.org/10.1021/nl5001899}.

\bibitem[{\citenamefont{Zang et~al.}(2011)\citenamefont{Zang, Mostovoy, Han,
  and Nagaosa}}]{zang_dynamics_2011}
\bibinfo{author}{\bibfnamefont{J.}~\bibnamefont{Zang}},
  \bibinfo{author}{\bibfnamefont{M.}~\bibnamefont{Mostovoy}},
  \bibinfo{author}{\bibfnamefont{J.~H.} \bibnamefont{Han}}, \bibnamefont{and}
  \bibinfo{author}{\bibfnamefont{N.}~\bibnamefont{Nagaosa}},
  \bibinfo{journal}{Physical Review Letters} \textbf{\bibinfo{volume}{107}},
  \bibinfo{pages}{136804} (\bibinfo{year}{2011}),
  \urlprefix\url{http://link.aps.org/doi/10.1103/PhysRevLett.107.136804}.

\bibitem[{\citenamefont{Schulz et~al.}(2012)\citenamefont{Schulz, Ritz, Bauer,
  Halder, Wagner, Franz, Pfleiderer, Everschor, Garst, and
  Rosch}}]{schulz_emergent_2012}
\bibinfo{author}{\bibfnamefont{T.}~\bibnamefont{Schulz}},
  \bibinfo{author}{\bibfnamefont{R.}~\bibnamefont{Ritz}},
  \bibinfo{author}{\bibfnamefont{A.}~\bibnamefont{Bauer}},
  \bibinfo{author}{\bibfnamefont{M.}~\bibnamefont{Halder}},
  \bibinfo{author}{\bibfnamefont{M.}~\bibnamefont{Wagner}},
  \bibinfo{author}{\bibfnamefont{C.}~\bibnamefont{Franz}},
  \bibinfo{author}{\bibfnamefont{C.}~\bibnamefont{Pfleiderer}},
  \bibinfo{author}{\bibfnamefont{K.}~\bibnamefont{Everschor}},
  \bibinfo{author}{\bibfnamefont{M.}~\bibnamefont{Garst}}, \bibnamefont{and}
  \bibinfo{author}{\bibfnamefont{A.}~\bibnamefont{Rosch}},
  \bibinfo{journal}{Nature Physics} \textbf{\bibinfo{volume}{8}},
  \bibinfo{pages}{301} (\bibinfo{year}{2012}), ISSN \bibinfo{issn}{1745-2473},
  \urlprefix\url{http://www.nature.com/nphys/journal/v8/n4/full/nphys2231.html%
}.

\bibitem[{\citenamefont{Neubauer et~al.}(2009)\citenamefont{Neubauer,
  Pfleiderer, Binz, Rosch, Ritz, Niklowitz, and
  B{\"o}ni}}]{neubauer_topological_2009}
\bibinfo{author}{\bibfnamefont{A.}~\bibnamefont{Neubauer}},
  \bibinfo{author}{\bibfnamefont{C.}~\bibnamefont{Pfleiderer}},
  \bibinfo{author}{\bibfnamefont{B.}~\bibnamefont{Binz}},
  \bibinfo{author}{\bibfnamefont{A.}~\bibnamefont{Rosch}},
  \bibinfo{author}{\bibfnamefont{R.}~\bibnamefont{Ritz}},
  \bibinfo{author}{\bibfnamefont{P.~G.} \bibnamefont{Niklowitz}},
  \bibnamefont{and} \bibinfo{author}{\bibfnamefont{P.}~\bibnamefont{B{\"o}ni}},
  \bibinfo{journal}{Physical Review Letters} \textbf{\bibinfo{volume}{102}},
  \bibinfo{pages}{186602} (\bibinfo{year}{2009}),
  \urlprefix\url{http://link.aps.org/doi/10.1103/PhysRevLett.102.186602}.

\bibitem[{\citenamefont{Lee et~al.}(2009)\citenamefont{Lee, Kang, Onose,
  Tokura, and Ong}}]{lee_unusual_2009}
\bibinfo{author}{\bibfnamefont{M.}~\bibnamefont{Lee}},
  \bibinfo{author}{\bibfnamefont{W.}~\bibnamefont{Kang}},
  \bibinfo{author}{\bibfnamefont{Y.}~\bibnamefont{Onose}},
  \bibinfo{author}{\bibfnamefont{Y.}~\bibnamefont{Tokura}}, \bibnamefont{and}
  \bibinfo{author}{\bibfnamefont{N.~P.} \bibnamefont{Ong}},
  \bibinfo{journal}{Physical Review Letters} \textbf{\bibinfo{volume}{102}},
  \bibinfo{pages}{186601} (\bibinfo{year}{2009}),
  \urlprefix\url{http://link.aps.org/doi/10.1103/PhysRevLett.102.186601}.

\bibitem[{\citenamefont{Kanazawa et~al.}(2011)\citenamefont{Kanazawa, Onose,
  Arima, Okuyama, Ohoyama, Wakimoto, Kakurai, Ishiwata, and
  Tokura}}]{kanazawa_large_2011}
\bibinfo{author}{\bibfnamefont{N.}~\bibnamefont{Kanazawa}},
  \bibinfo{author}{\bibfnamefont{Y.}~\bibnamefont{Onose}},
  \bibinfo{author}{\bibfnamefont{T.}~\bibnamefont{Arima}},
  \bibinfo{author}{\bibfnamefont{D.}~\bibnamefont{Okuyama}},
  \bibinfo{author}{\bibfnamefont{K.}~\bibnamefont{Ohoyama}},
  \bibinfo{author}{\bibfnamefont{S.}~\bibnamefont{Wakimoto}},
  \bibinfo{author}{\bibfnamefont{K.}~\bibnamefont{Kakurai}},
  \bibinfo{author}{\bibfnamefont{S.}~\bibnamefont{Ishiwata}}, \bibnamefont{and}
  \bibinfo{author}{\bibfnamefont{Y.}~\bibnamefont{Tokura}},
  \bibinfo{journal}{Phys. Rev. Lett.} \textbf{\bibinfo{volume}{106}},
  \bibinfo{pages}{156603} (\bibinfo{year}{2011}),
  \urlprefix\url{http://link.aps.org/doi/10.1103/PhysRevLett.106.156603}.

\bibitem[{\citenamefont{Li et~al.}(2013)\citenamefont{Li, Kanazawa, Yu,
  Tsukazaki, Kawasaki, Ichikawa, Jin, Kagawa, and Tokura}}]{li_robust_2013}
\bibinfo{author}{\bibfnamefont{Y.}~\bibnamefont{Li}},
  \bibinfo{author}{\bibfnamefont{N.}~\bibnamefont{Kanazawa}},
  \bibinfo{author}{\bibfnamefont{X.~Z.} \bibnamefont{Yu}},
  \bibinfo{author}{\bibfnamefont{A.}~\bibnamefont{Tsukazaki}},
  \bibinfo{author}{\bibfnamefont{M.}~\bibnamefont{Kawasaki}},
  \bibinfo{author}{\bibfnamefont{M.}~\bibnamefont{Ichikawa}},
  \bibinfo{author}{\bibfnamefont{X.~F.} \bibnamefont{Jin}},
  \bibinfo{author}{\bibfnamefont{F.}~\bibnamefont{Kagawa}}, \bibnamefont{and}
  \bibinfo{author}{\bibfnamefont{Y.}~\bibnamefont{Tokura}},
  \bibinfo{journal}{Physical Review Letters} \textbf{\bibinfo{volume}{110}},
  \bibinfo{pages}{117202} (\bibinfo{year}{2013}),
  \urlprefix\url{http://link.aps.org/doi/10.1103/PhysRevLett.110.117202}.

\bibitem[{\citenamefont{Watanabe et~al.}(2014)\citenamefont{Watanabe,
  Parameswaran, Raghu, and Vishwanath}}]{watanabe_anomalous_2014}
\bibinfo{author}{\bibfnamefont{H.}~\bibnamefont{Watanabe}},
  \bibinfo{author}{\bibfnamefont{S.~A.} \bibnamefont{Parameswaran}},
  \bibinfo{author}{\bibfnamefont{S.}~\bibnamefont{Raghu}}, \bibnamefont{and}
  \bibinfo{author}{\bibfnamefont{A.}~\bibnamefont{Vishwanath}},
  \bibinfo{journal}{Physical Review B} \textbf{\bibinfo{volume}{90}},
  \bibinfo{pages}{045145} (\bibinfo{year}{2014}),
  \urlprefix\url{http://link.aps.org/doi/10.1103/PhysRevB.90.045145}.

\bibitem[{\citenamefont{Huang et~al.}(2014{\natexlab{a}})\citenamefont{Huang,
  Chen, Kang, Zang, Shu, Chou, and Chien}}]{huang_magnetoresistance_2014}
\bibinfo{author}{\bibfnamefont{S.~X.} \bibnamefont{Huang}},
  \bibinfo{author}{\bibfnamefont{F.}~\bibnamefont{Chen}},
  \bibinfo{author}{\bibfnamefont{J.}~\bibnamefont{Kang}},
  \bibinfo{author}{\bibfnamefont{J.}~\bibnamefont{Zang}},
  \bibinfo{author}{\bibfnamefont{G.~J.} \bibnamefont{Shu}},
  \bibinfo{author}{\bibfnamefont{F.~C.} \bibnamefont{Chou}}, \bibnamefont{and}
  \bibinfo{author}{\bibfnamefont{C.~L.} \bibnamefont{Chien}},
  \bibinfo{journal}{{arXiv}:1409.7867 [cond-mat]}
  (\bibinfo{year}{2014}{\natexlab{a}}), \bibinfo{note}{{arXiv}: 1409.7867},
  \urlprefix\url{http://arxiv.org/abs/1409.7867}.

\bibitem[{\citenamefont{Huang et~al.}(2014{\natexlab{b}})\citenamefont{Huang,
  Kang, Chen, Zang, Shu, Chou, Grigoriev, Dyadkin, and
  Chien}}]{huang_universal_2014}
\bibinfo{author}{\bibfnamefont{S.~X.} \bibnamefont{Huang}},
  \bibinfo{author}{\bibfnamefont{J.}~\bibnamefont{Kang}},
  \bibinfo{author}{\bibfnamefont{F.}~\bibnamefont{Chen}},
  \bibinfo{author}{\bibfnamefont{J.}~\bibnamefont{Zang}},
  \bibinfo{author}{\bibfnamefont{G.~J.} \bibnamefont{Shu}},
  \bibinfo{author}{\bibfnamefont{F.~C.} \bibnamefont{Chou}},
  \bibinfo{author}{\bibfnamefont{S.~V.} \bibnamefont{Grigoriev}},
  \bibinfo{author}{\bibfnamefont{V.~A.} \bibnamefont{Dyadkin}},
  \bibnamefont{and} \bibinfo{author}{\bibfnamefont{C.~L.} \bibnamefont{Chien}},
  \bibinfo{journal}{{arXiv}:1409.7869 [cond-mat]}
  (\bibinfo{year}{2014}{\natexlab{b}}), \bibinfo{note}{{arXiv}: 1409.7869},
  \urlprefix\url{http://arxiv.org/abs/1409.7869}.

\bibitem[{\citenamefont{Winkler}(2003)}]{winkler_spin--orbit_2003}
\bibinfo{author}{\bibfnamefont{R.}~\bibnamefont{Winkler}},
  \emph{\bibinfo{title}{Spin--Orbit Coupling Effects in Two-Dimensional
  Electron and Hole Systems}}, Springer Tracts in Modern Physics
  (\bibinfo{publisher}{Springer}, \bibinfo{year}{2003}).

\bibitem[{\citenamefont{Dresselhaus et~al.}(2010)\citenamefont{Dresselhaus,
  Dresselhaus, and Jorio}}]{dresselhaus_group_2010}
\bibinfo{author}{\bibfnamefont{M.~S.} \bibnamefont{Dresselhaus}},
  \bibinfo{author}{\bibfnamefont{G.}~\bibnamefont{Dresselhaus}},
  \bibnamefont{and} \bibinfo{author}{\bibfnamefont{A.}~\bibnamefont{Jorio}},
  \emph{\bibinfo{title}{Group Theory: Application to the Physics of Condensed
  Matter}} (\bibinfo{publisher}{Springer}, \bibinfo{address}{Berlin},
  \bibinfo{year}{2010}), \bibinfo{edition}{softcover reprint of hardcover 1st
  ed. 2008 edition} ed., ISBN \bibinfo{isbn}{9783642069451}.

\bibitem[{\citenamefont{Jeong and Pickett}(2004)}]{jeong_implications_2004}
\bibinfo{author}{\bibfnamefont{T.}~\bibnamefont{Jeong}} \bibnamefont{and}
  \bibinfo{author}{\bibfnamefont{W.}~\bibnamefont{Pickett}},
  \bibinfo{journal}{Physical Review B} \textbf{\bibinfo{volume}{70}},
  \bibinfo{pages}{075114} (\bibinfo{year}{2004}),
  \urlprefix\url{http://link.aps.org/doi/10.1103/PhysRevB.70.075114}.

\bibitem[{\citenamefont{Koster}(1957)}]{koster_space_1957}
\bibinfo{author}{\bibfnamefont{G.~F.} \bibnamefont{Koster}},
  \emph{\bibinfo{title}{Space Groups and Their Representations}}
  (\bibinfo{publisher}{Academic Press}, \bibinfo{year}{1957}), ISBN
  \bibinfo{isbn}{9780124337848}.

\bibitem[{\citenamefont{{\v Z}uti{\'c} et~al.}(2004)\citenamefont{{\v
  Z}uti{\'c}, Fabian, and Das~Sarma}}]{zutic_spintronics:_2004}
\bibinfo{author}{\bibfnamefont{I.}~\bibnamefont{{\v Z}uti{\'c}}},
  \bibinfo{author}{\bibfnamefont{J.}~\bibnamefont{Fabian}}, \bibnamefont{and}
  \bibinfo{author}{\bibfnamefont{S.}~\bibnamefont{Das~Sarma}},
  \bibinfo{journal}{Reviews of Modern Physics} \textbf{\bibinfo{volume}{76}},
  \bibinfo{pages}{323} (\bibinfo{year}{2004}),
  \urlprefix\url{http://link.aps.org/doi/10.1103/RevModPhys.76.323}.

\bibitem[{\citenamefont{Leonov}(2014)}]{leonov_chiral_2014}
\bibinfo{author}{\bibfnamefont{A.~O.} \bibnamefont{Leonov}},
  \bibinfo{journal}{{arXiv}:1406.2177 [cond-mat]}  (\bibinfo{year}{2014}),
  \bibinfo{note}{{arXiv}: 1406.2177},
  \urlprefix\url{http://arxiv.org/abs/1406.2177}.

\bibitem[{\citenamefont{Ruderman and Kittel}(1954)}]{ruderman_indirect_1954}
\bibinfo{author}{\bibfnamefont{M.~A.} \bibnamefont{Ruderman}} \bibnamefont{and}
  \bibinfo{author}{\bibfnamefont{C.}~\bibnamefont{Kittel}},
  \bibinfo{journal}{Physical Review} \textbf{\bibinfo{volume}{96}},
  \bibinfo{pages}{99} (\bibinfo{year}{1954}),
  \urlprefix\url{http://link.aps.org/doi/10.1103/PhysRev.96.99}.

\bibitem[{\citenamefont{Kasuya}(1956)}]{kasuya_theory_1956}
\bibinfo{author}{\bibfnamefont{T.}~\bibnamefont{Kasuya}},
  \bibinfo{journal}{Progress of Theoretical Physics}
  \textbf{\bibinfo{volume}{16}}, \bibinfo{pages}{45} (\bibinfo{year}{1956}),
  ISSN \bibinfo{issn}{0033-068X, 1347-4081},
  \urlprefix\url{http://ptp.oxfordjournals.org/content/16/1/45}.

\bibitem[{\citenamefont{Yosida}(1957)}]{yosida_magnetic_1957}
\bibinfo{author}{\bibfnamefont{K.}~\bibnamefont{Yosida}},
  \bibinfo{journal}{Physical Review} \textbf{\bibinfo{volume}{106}},
  \bibinfo{pages}{893} (\bibinfo{year}{1957}),
  \urlprefix\url{http://link.aps.org/doi/10.1103/PhysRev.106.893}.

\bibitem[{\citenamefont{Fert et~al.}(2013)\citenamefont{Fert, Cros, and
  Sampaio}}]{fert_skyrmions_2013}
\bibinfo{author}{\bibfnamefont{A.}~\bibnamefont{Fert}},
  \bibinfo{author}{\bibfnamefont{V.}~\bibnamefont{Cros}}, \bibnamefont{and}
  \bibinfo{author}{\bibfnamefont{J.}~\bibnamefont{Sampaio}},
  \bibinfo{journal}{Nature nanotechnology} \textbf{\bibinfo{volume}{8}},
  \bibinfo{pages}{152} (\bibinfo{year}{2013}),
  \urlprefix\url{http://dx.doi.org/10.1038/nnano.2013.29}.

\bibitem[{\citenamefont{Bazaliy et~al.}(1998)\citenamefont{Bazaliy, Jones, and
  Zhang}}]{bazaliy_modification_1998}
\bibinfo{author}{\bibfnamefont{Y.~B.} \bibnamefont{Bazaliy}},
  \bibinfo{author}{\bibfnamefont{B.~A.} \bibnamefont{Jones}}, \bibnamefont{and}
  \bibinfo{author}{\bibfnamefont{S.-C.} \bibnamefont{Zhang}},
  \bibinfo{journal}{Physical Review B} \textbf{\bibinfo{volume}{57}},
  \bibinfo{pages}{R3213} (\bibinfo{year}{1998}),
  \urlprefix\url{http://link.aps.org/doi/10.1103/PhysRevB.57.R3213}.

\bibitem[{\citenamefont{Tatara and Kohno}(2004)}]{tatara_theory_2004}
\bibinfo{author}{\bibfnamefont{G.}~\bibnamefont{Tatara}} \bibnamefont{and}
  \bibinfo{author}{\bibfnamefont{H.}~\bibnamefont{Kohno}},
  \bibinfo{journal}{Physical Review Letters} \textbf{\bibinfo{volume}{92}},
  \bibinfo{pages}{086601} (\bibinfo{year}{2004}),
  \urlprefix\url{http://link.aps.org/doi/10.1103/PhysRevLett.92.086601}.

\bibitem[{\citenamefont{Bauer and Pfleiderer}(2012)}]{bauer_magnetic_2012}
\bibinfo{author}{\bibfnamefont{A.}~\bibnamefont{Bauer}} \bibnamefont{and}
  \bibinfo{author}{\bibfnamefont{C.}~\bibnamefont{Pfleiderer}},
  \bibinfo{journal}{Physical Review B} \textbf{\bibinfo{volume}{85}},
  \bibinfo{pages}{214418} (\bibinfo{year}{2012}),
  \urlprefix\url{http://link.aps.org/doi/10.1103/PhysRevB.85.214418}.

\bibitem[{\citenamefont{Levy and Zhang}(1997)}]{levy_resistivity_1997}
\bibinfo{author}{\bibfnamefont{P.~M.} \bibnamefont{Levy}} \bibnamefont{and}
  \bibinfo{author}{\bibfnamefont{S.}~\bibnamefont{Zhang}},
  \bibinfo{journal}{Physical Review Letters} \textbf{\bibinfo{volume}{79}},
  \bibinfo{pages}{5110} (\bibinfo{year}{1997}),
  \urlprefix\url{http://link.aps.org/doi/10.1103/PhysRevLett.79.5110}.

\end{thebibliography}

\begin{thebibliography}{10}
  \bibitem{DomainWall} P. M. Levy, and S. Zhang, Phys. Rev. Lett. \textbf{79}, 5110 (1997).
  \bibitem{BZShort}  R. Hlubina and T. M. Rice, Phys. Rev. B. \textbf{51}, 9253 (1995).
  \bibitem{Ziman} Ziman, Electrons and Phonons (Clarendon, Oxford, 1960).
\end{thebibliography}
\end{document}